\documentclass[journal]{IEEEtran}

\usepackage{caption}
\usepackage{subcaption}
\usepackage{float}
\usepackage[T1]{fontenc}
\usepackage[latin1]{inputenc} 
\usepackage{babel}
\usepackage[font=small,labelfont=bf]{caption}
\DeclareCaptionLabelFormat{andtable}{#1~#2  \&  \tablename~\thetable}

\usepackage{graphicx}
\usepackage{array}
\usepackage{epsfig}
\usepackage{epstopdf}
\usepackage{latexsym}
\usepackage{amsfonts}
\usepackage{here}
\usepackage{rawfonts}
\usepackage{calc}
\usepackage{url}
\usepackage{enumerate}
\usepackage{color}
\usepackage[tbtags]{amsmath}
\usepackage{amssymb}
\usepackage{amsmath}
\usepackage{upref}
\usepackage{times}
\usepackage{array}

\usepackage[noadjust]{cite}

\newtheorem{theorem}{Theorem}
\newtheorem{lemma}{Lemma}
\newtheorem{definition}{Definition}
\newtheorem{proposition}{Proposition}

\begin{document}
	
	\title{Do Not Forget the Past: A Buffer-Aided Framework for Relay Based Key Generation}
	
	\author{Rusni Kima Mangang and J. Harshan\\
		Indian Institute of Technology Delhi, India.
	}
	
	\maketitle
	
\begin{abstract}
We address relay-assisted key generation wherein two wireless nodes, that have no direct channel between them, seek the assistance of an intermediate relay to generate secret keys. In a celebrated version of the relay-assisted protocol, as applied by \emph{Lai et al.}, \emph{Zhou et al.}, \emph{Wang et al.} and \emph{Waqas et al.}, the relay node generates pair-wise keys with the two nodes, and then broadcasts an XOR version of the two keys. Although such protocols are simple and effective, we observe that they face reduction in key rates due to two problems. First, for confidentiality, the relay broadcasts an XOR function of the pair-wise keys thereby pruning the length of the shared key to be the minimum of the key lengths of the pair-wise keys. Secondly, the broadcast phase may also experience outages thereby not being able to share the generated key in every round of the protocol. Identifying these issues, we propose a buffer-aided relaying protocol wherein buffer is used at the relay to store unused secret bits generated in the previous rounds of the protocol so as to provide confidentiality in the subsequent rounds of broadcast. On this buffer-aided protocol, we propose a power-allocation strategy between the phases of key generation and broadcast so as to maximize the throughput and key rate. Rigorous analyses show that buffer-aided relay when implemented along with the proposed power-allocation strategy offer remarkable advantages over existing baselines.
 

		%
	\end{abstract}
	
\section{Introduction}
	
It is well known that eavesdropping over wireless channels can be mitigated through symmetric-key crypto-primitives \cite{ri}, \cite{umm}. Although crypto-primitive based techniques are effective in providing confidentiality, they necessitate the participating nodes to synthesize shared secret-keys at regular intervals. In the context of standard cellular communication, say between a user equipment and a base-station, secret-keys are synthesized using pre-registered subscriber identity module (SIM) based information between the two entities. However, in the context of device-to-device communication, wherein no pre-registered SIM based informations are available between the devices, additional dynamic key-generation techniques must be implemented. Among several such key-generation techniques, physical-layer key (PLK) generation methods have received traction in the wireless community. Specifically, in PLK generation, two wireless nodes exploit random fluctuations in their wireless channels to synthesize shared secret-keys, and subsequently feed the generated keys to the higher-level crypto-blocks \cite{adw}-\cite{LLP}. 

\subsection{Motivation}
\label{subsec:motivation}

In this work, we are interested in developing new strategies for PLK generation particularly addressing practical issues that forbid key generation. For instance, PLK generation is not feasible when the two devices are out-of-coverage due to signal-blockage issues or power-limitations. One possible direction to circumvent this problem is to take the assistance of a trust-worthy neighboring node \cite{lyw}-\cite{qkk} that acts as a relay by offering the much needed connectivity and a random channel for the two nodes. To formally explain the relay based PLK generation, consider a wireless communication setting between two nodes, denoted by Node-A and Node-B, as shown in Fig. \ref{fig:network_model}, which would like to harvest shared secret-keys through PLK generation. When the
direct channel between Node-A and Node-B is not available, the two nodes use a trusted relay, denoted by Node-R, that can offer wireless channels with significant randomness. With the help of the relay, there are broadly two options for key generation; firstly, the relay can share the common randomness with the legitimate nodes through amplify-and-forward strategy \cite{SIS},\cite{TLQ}. Secondly, the relay can generate keys with the nodes and broadcast them confidentially \cite{lyw},\cite{hlll}. However, among the two options, the amplify-and-forward strategy is not preferred due to higher noise in the common source of randomness as well as loss of key rate due to leakage issues in the common randomness at an eavesdropper. With the help of the relay under the latter class, it is clear that Node-A and Node-R may harvest keys by exchanging pilot symbols during the so-called \emph{key generation phase}. Subsequently, this generated key must be reliably shared with Node-B during a dedicated time called the \emph{broadcast phase}. Note that Node-R must confidentially share the generated key with Node-B, and this implies that the broadcast phase also requires a pre-shared key between Node-R and Node-B. Overall, the relay based PLK generation is such that pair-wise keys must be generated between Node-A and Node-R, and Node-B and Node-R, in the key generation phase, and then, one of the keys is used to securely share the other key in the broadcast phase. 

Among the existing relay based PLK generation methods, a state-of-the-art technique \cite{lyw},\cite{hlll},\cite{qkk},\cite{waljin} that has gathered attention due to its simple and effective implementation, works as follows: \textbf{Step 1:} Pair-wise secret-keys are generated between Node-A and Node-R, denoted by $k_{AR} \in \{0, 1\}^{N_{AR}}$, and Node-B and Node-R, denoted by $k_{BR} \in \{0, 1\}^{N_{BR}}$ during the key generation phase, where $N_{AR}$ and $N_{BR}$ denote the lengths of the keys.  \textbf{Step 2:} In the broadcast phase, an XOR version of pair-wise keys is shared by Node-R so that Node-B can retrieve $k_{AR}$ through self-interference cancellation since $k_{BR}$ is known at Node-B. For this celebrated XOR based idea, we identify two problems: 

	\begin{figure*}
		\begin{center}
			\includegraphics[width=6in, height=2.8in]{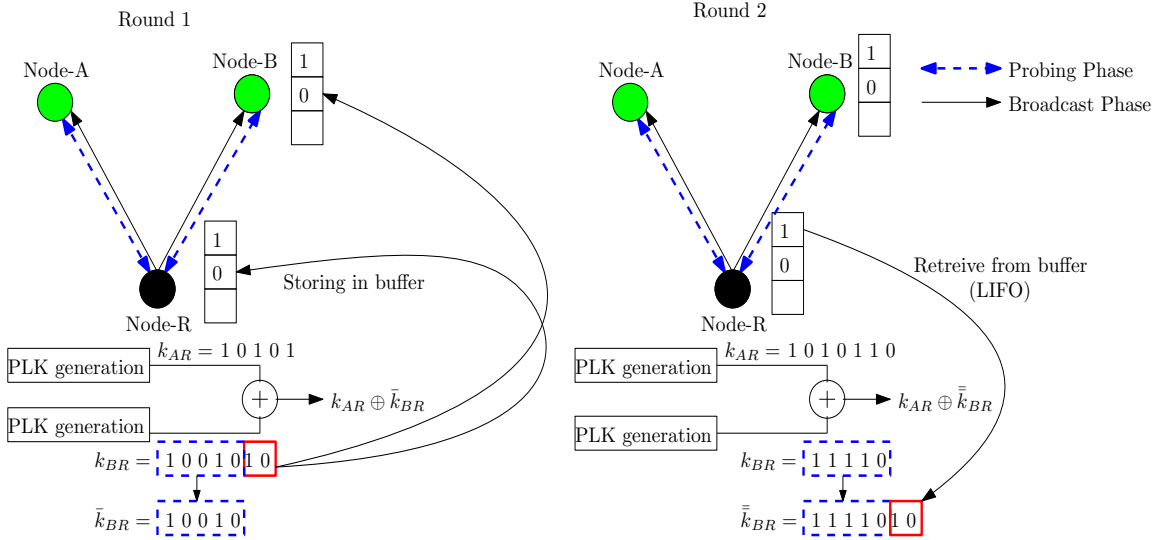}
		\end{center}
		\linespread{1}\selectfont{}
		\caption{\label{fig:network_model} Network model comprising Node-A and Node-B which intend to harvest secret-keys using their channel with Node-R. We use buffer-aided relay along with a power-allocation strategy between the key generation phase and broadcast phase to improve the throughput and key rates of the protocol. Observe that the unused secret bits of Round 1 are used in Round 2.}
	\end{figure*}

\noindent \textbf{Problem 1}: Although Node-A and Node-R, and Node-B and Node-R generate $k_{AR}$ and $k_{BR}$ in the key generation phase, respectively, the length of the key that is shared with Node-B in the broadcast phase is min($N_{AR},N_{BR}$), which in turn reduces the overall key-rate of the protocol. Note that this limitation arises because of the XOR operation that provides confidentiality from an external eavesdropper listening to the broadcast phase. Motivated by this problem, in this paper, we ask \emph{``How do we circumvent this loss in key-rate due to the XOR operation?"} Given that key-generation protocols are typically executed in multiple rounds in order to update the secret-keys, we explore the possibility of using buffer-aided relay to answer the above question.

\noindent \textbf{Problem 2}: We note that the randomness offered by the wireless channel between Node-A (or Node-B) and Node-R depends on its Line-of-Sight (LOS) component. In other words, higher is the LOS component of the channel, lower is the key length, and vice-versa. On the other hand, since the generated key must be broadcast through the same channel characteristics, we note that higher the LOS component of the channel, higher the reliability. Thus, a lower LOS channel offers conflicting behavior during the key generation phase and the broadcast phase. Moreover, with a given total power budget, allocating majority fraction of power to the key generation phase results in high-rate pair-wise keys; however, this leads to
outage events owing to insufficient power to deliver those keys during the broadcast phase. On the
other hand, allocating minority fraction of power to probing signals results in low-rate pair-wise secret-keys; however, although this may reduce the fraction of outage events owing to significant power on the
broadcast signal, the overall key-rate at Node-A and Node-B might not be maximized. With this observation, we ask \emph{``For a given LOS characteristics of the channel, how should the three nodes distribute their power between the key generation and the broadcast phase so as to maximize number of secret bits that reach Node-B?"}

\subsection{Contributions}
	
\noindent (1) We propose buffer-aided relay for PLK generation, and then study its impact on the overall key rate of the protocol. With the buffer at the relay, we show that unused secret bits generated between Node-R and Node-B can be temporarily stored at the relay, and these bits could be used to provide confidentiality  for the broadcast phase in the subsequent rounds of key generation. This way, the message length in the broadcast phase would be more than the minimum of the key lengths generated between Node-A and Node-R, and Node-B and Node-R, thereby improving the key rate (see Section \ref{sec:relay_model}).  

\noindent (2) When using buffer-aided relay we study power-allocation strategies between the key generation phase and the broadcast phase such that the number of secret bits generated between Node-A and Node-B is maximized as a function of the LOS parameter between Node-A (or Node-B) and Node-R, and the underlying signal-to-noise-ratio (SNR). We first take up the power-allocation problem when optimal key generation algorithms are employed at the three nodes
(see Section \ref{sec:asym_throuput}). Subsequently, we formulate optimization problems to (i) maximize the \emph{throughput} of the protocol, and to (ii) maximize the \emph{key rate} of the protocol subject to an upper bound on the error-rate, say some $\eta > 0$, of the broadcast phase. The \emph{key rate} metric is useful when the higher-level crypto-primitives of Node-A and Node-B expect shared secret bits through Node-R on at least $1- \eta$ fraction of the key generation rounds. The number $\eta$ is however chosen such that the two nodes can manage to garner shared secret bits from some other key generation methods on the residual $\eta$ fraction. In a different scenario, the \emph{throughput} metric is useful when the crypto-primitives of Node-A and Node-B do not impose strict requirements to generate the shared secret bits as long as the total number of bits generated across several key generation rounds is maximized. We present an extensive analysis on the objective functions and the underlying constraints of the optimization problem to show that standard gradient-descent algorithms can be applied to obtain near-optimal solutions. Through simulation results, we show that the proposed solutions provide significant benefits over standard baselines.
		
\noindent (3) We also extend the power-allocation problem to practical scenarios wherein (i) practical key generation algorithms are employed at the three nodes, and (ii) the buffer size at the beginning of the protocol is empty (see Section \ref{sec:prac_constraints}). In this case, we observe that the key rates offered by the protocol is a correlated process since the buffer also accumulates shared secret bits with Node-B across successive rounds of the key generation protocol. As a result, we capture the update process of the buffer as a function of successive rounds, and then derive closed form expressions on key rate. Finally, owing to short block-length codes for the broadcast phase, we invoke non-asymptotic outage probability results from \cite{yuri_FBL}, to pose an optimization problem over the power-allocation variable. Through simulation results, we show that the optimal power-allocation parameter results in substantial benefits in both throughput and key rate over equal power allocation between the key generation phase and the broadcast phase. Finally, we show that the key generation protocol with buffer-aided relay outperforms the one without buffer \cite{lyw},\cite{hlll},\cite{qkk},\cite{waljin}. 

\noindent (4) Finally, we discuss the application of buffer-aided protocol on relay networks wherein the LOS components between the nodes and the relay are different. We show that the optimization problems discussed in the context of relay networks with equal LOS components continue to apply for the case of unequal LOS components (see Section \ref{sec:unequal_c}).

%

\begin{figure}[H]
	\centering
	\begin{subfigure}{0.65\linewidth}
		\includegraphics[scale = 1]{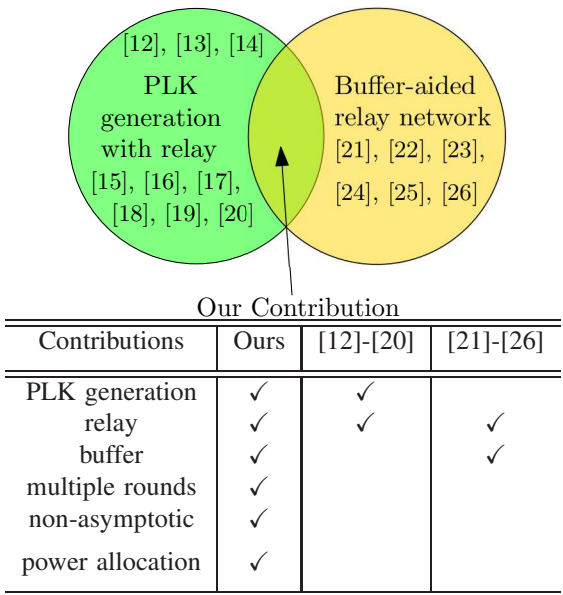}
	\end{subfigure}
	\begin{subfigure}{1\linewidth}
			\centering
			\begin{tabular}[b]{c|c|c|c} 
					\hline\hline 
					Contributions&Ours&\cite{lyw}-\cite{NAHM}&\cite{ja_secrecy_throughput}-\cite{JJYX_Norio_security_delay}\\ [1ex] 
					\hline \hline 
					PLK generation & \checkmark & \checkmark & \\ 
					relay & \checkmark & \checkmark & \checkmark\\
					buffer & \checkmark &  &\checkmark\\
					multiple rounds & \checkmark &  & \\
					non-asymptotic & \checkmark &  &\\ [1ex] 
					power allocation & \checkmark &  &\\ [1ex]
					\hline 
			\end{tabular}%
	\end{subfigure}
	\caption{Salient features of our work: (i) multiple rounds of PLK generation, (ii) buffers, and (iii) power-allocation strategy.}
	\label{fig:contribution_venn}
\end{figure}


\subsection{Related Work and Novelty}

In Fig. \ref{fig:contribution_venn}, we have shown the novel contributions of our work in contrast to the existing contributions. The idea of using a trusted relay \cite{icpn} for key generation is closely related to the contributions of \cite{lyw},\cite{hlll}. However, unlike \cite{lyw} and \cite{hlll}, our work is different in the following aspects: (i) the relay is trusted\footnote{Given that Node-A and Node-B are out of coverage, it is implicit that a relay node would also be needed to forward the payload to Node-B. Thus, the use of trusted relay is natural to the setting. However, as a future direction for research, it would still be interesting to study buffer-aided relaying protocol along with an untrusted relay \cite{XDDK},\cite{GLMC},\cite{TLQ},\cite{NAHM}.} and equipped with a buffer thereby aiding to increase the key rate for every round of key generation, (ii) variable power is considered between the key generation phase and the broadcast phase to deliver maximum number of secret bits within every round of key generation, (iii) the proposed power-allocation strategy is applicable when the two channels are characterized by arbitrary degree of LOS components. As far as the use of buffers is concerned, existing relay based PLK generation methods have not considered buffers in their model. However, buffer-aided relays have been used to improve secrecy throughput \cite{ja_secrecy_throughput},\cite{jing_secrecy_throughput}, minimize secrecy outage probability \cite{RyoShi_secrecy_outage},\cite{xtyyh_secrecy_outage} or investigate security and delay/QoS trade-off in the presence of eavesdropper \cite{XYZY_security_delay},\cite{JJYX_Norio_security_delay}.  

\begin{figure}
	\begin{center}
		\includegraphics[scale= 1.2]{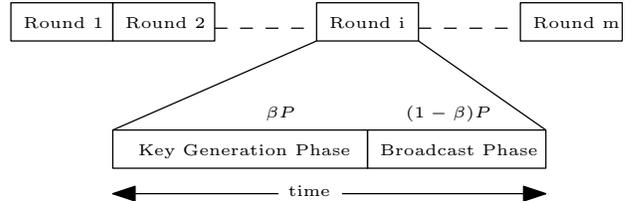}
	\end{center}
	\linespread{1}\selectfont{}
	\caption{\label{fig:phase_model} A single round of key sharing comprises two phases: the key generation phase and broadcast phase. The optimization parameter $\beta \in (0, 1)$ is used to allocate power between the two phases.}
\end{figure}


\section{Buffer-Aided Protocol with relay}
\label{sec:relay_model}

Consider a network model, as shown in Fig. \ref{fig:network_model}, wherein Node-A and Node-B take the assistance of Node-R to generate secret-keys using PLK generation framework. The complex base band channel between Node-A and Node-R is denoted by $h_{AR} = \sqrt{c_{R}}\left(\frac{1 + i}{\sqrt{2}}\right) + \sqrt{1 - c_{R}}g_{AR},$ such that the constant $\sqrt{c_{R}}\left(\frac{1 + i}{\sqrt{2}}\right)$ captures the LOS component with $c_{R} \in [0, 1]$ and $\sqrt{1 - c_{R}}g_{AR}$ captures the Non-LOS component with $g_{AR} \sim \mathcal{CN}(0, 1)$. As a special case, when $c_{R} = 0$ and $c_{R} = 1$, $h_{AR}$ corresponds to a Rayleigh channel and an Additive White Gaussian Noise (AWGN) channel, respectively.\footnote{In classical Ricean fading models \cite{K_factor}, the parameter $K \in [0, \infty]$ is used to capture various degrees of LOS components. Alternatively, in this model, we use $c_{R}$ for the same purpose, where $K$ can be obtained as $K = \frac{c_{R}}{1 - c_{R}}$.} On the other hand, when $c_{R} \in (0, 1)$, $h_{AR}$ corresponds to a Ricean channel with arbitrary degree of LOS as determined by $c_{R}$. Along the similar lines, the complex base band channel between Node-B and Node-R is denoted by $h_{BR} = \sqrt{c_{R}}\left(\frac{1 + i}{\sqrt{2}}\right) + \sqrt{1 - c_{R}}g_{BR},$ such that $\sqrt{1 - c_{R}}g_{BR}$ captures the Non-LOS component with $g_{BR} \sim \mathcal{CN}(0, 1)$. We assume that both $h_{AR}$ and $h_{BR}$ are quasi-static with a coherence interval of $T$ time-slots, henceforth referred to as a coherence-block. Although identical LOS components are assumed for the two channels, note that their channel realizations are independent owing to their statistically independent Non-LOS components. Along the similar lines of \cite{wts_reciprocity},\cite{rndhl_reciprocity},\cite{hlll},\cite{waljin}, we assume perfect reciprocity in the pair-wise channels.\footnote{Pair-wise key generation protocol is still applicable for imperfect channel reciprocity \cite{qkk},\cite{fha_imperfect}. Although the key-rate reduces due to imperfect reciprocity, the difference in the key length of the two channels persists, and therefore, buffers are applicable.} In order to assist buffer-aided key generation, we assume that Node-B and Node-R are equipped with a buffer, using which multiple rounds of relay-assisted key generation protocols are executed, as shown in Fig. \ref{fig:phase_model}. To maintain consistency, the two nodes use the buffer as a stack with Last-In First-Out (LIFO) protocol. To capture successive rounds of key-generation protocols, we use $m \geq 0$ to denote the round number, and then explain the key-generation protocol for the $m$-th round in the next section.

\subsection{Protocol for Key Generation and Distribution}
\label{buffer_protocol}

To explain the buffer-aided key generation protocol on the $m$-th round, we assume that the buffers at Node-R  and Node-B contain $B(m-1) \geq 0$ number of unused secret bits that were generated from the previous rounds. As shown in Fig. \ref{fig:phase_model}, each round of key generation protocol constitutes $L+1$ coherence-blocks, with $T \ge L$ for some $L \in \mathbb{Z}_{+}$, within which (i) $L$ coherence-blocks are used by the three nodes for the key generation phase to generate pair-wise keys, and (ii) the $(L+1)$-th coherence-block is used by Node-R to broadcast the generated key to Node-B. A total power budget of $PL$ units is divided between the key generation phase and the broadcast phase as $\beta P L$ and $(1 - \beta) P L$, for some $0 < \beta < 1$. Subsequently, $\beta P L$ is equally divided among the three nodes over $L$ coherence-blocks for transmitting the probing symbols, and $(1 - \beta) P L$ is used by Node-R to distribute the generated keys to Node-B during the $(L+1)$-th coherence-block. This way, we use $\beta$ as the underlying optimization parameter to distribute the power between the key generation phase and the broadcast phase. Note that this framework of power-allocation requires the total power at the relay to be higher than that of the other nodes since it has to execute both the probing phase and the broadcast phase. We do not consider power-allocation for the consensus-phase of key generation as it is proportional to the power needed for probing signals.
By using $l \in \{1, 2, \ldots, L + 1\}$ to denote the coherence-block index, the two channels in the network are represented as $h_{AR}(l)$ and $h_{BR}(l)$. To keep the notations simple, we do not use the round index $m$ in the signal model, however, we use it when referring to the key lengths and the buffer size. First, we present a description of the key generation phase of the protocol within a coherence-block index $l$, for $1 \leq l \leq L$, and then describe the broadcast phase in the $(L+1)$-th coherence-block.


\subsubsection{Key Generation Phase}

In the key generation phase, Node-A, Node-B and Node-R take turns to broadcast a  symbol $x = \sqrt{\frac{\beta P}{3}}$ during the $1$st, $2$nd and $3$rd slots of the $l$-th coherence-block.\\ \textbf{Slot 1}: Node-A transmits a probing signal $x = \sqrt{\frac{\beta P}{3}}$, which is used by Node-R to receive a noisy version of $h_{AR}(l)$. In particular, the corresponding received signal is
\begin{equation}
\label{eq:p2_eq1}
y^{(1)}_{R}(l) = \sqrt{\frac{\beta P}{3}} h_{AR}(l) + n^{(1)}_{R}(l),
\end{equation}
where $n^{(1)}_{R}(l)$ represents the AWGN, distributed as $\mathcal{CN}(0, \gamma)$. Here, the superscript denotes the slot number of the coherence-block. 

\noindent \textbf{Slot 2}: Node-B transmits a probing signal $x = \sqrt{\frac{\beta P}{3}}$, which is used by Node-R to receive a noisy version of $h_{BR}(l)$. In particular, the corresponding received signal is
\begin{equation}
\label{eq:p2_eq1}
y^{(2)}_{R}(l) = \sqrt{\frac{\beta P}{3}} h_{BR}(l) + n^{(2)}_{R}(l),
\end{equation}
where $n^{(2)}_{R}(l)$ represents the AWGN, distributed as $\mathcal{CN}(0, \gamma)$.

\noindent \textbf{Slot 3}: Node-R transmits a probing signal $x = \sqrt{\frac{\beta P}{3}}$, which is used by Node-A and Node-B to receive a noisy version of $h_{AR}(l)$ and $h_{BR}(l)$, respectively. In particular, the corresponding received signals are

\begin{equation}
\label{eq:p3_eq1}
\resizebox{\columnwidth}{!} 
{
$y^{(3)}_{A}(l) = \sqrt{\frac{\beta P}{3}} h_{AR}(l) + n^{(3)}_{A}(l) \mbox{ \& }y^{(3)}_{B}(l) = \sqrt{\frac{\beta P}{3}} h_{BR}(l) + n^{(3)}_{B}(l)$,
}
\end{equation}


where $n^{(3)}_{A}(l)$ and $n^{(3)}_{B}(l)$ represent the AWGN, distributed as $\mathcal{CN}(0, \gamma)$.  Using the first $L$ coherence-blocks of the $m$-th round, Node-A and Node-R use their observations $\{y^{(1)}_{R}(l), 1 \leq l \leq L\}$ and $\{y^{(3)}_{A}(l), 1 \leq l \leq L\}$, respectively, to apply a pair-wise key generation algorithm to synthesize $N_{AR}(m)$ secret bits, denoted by $k_{AR}(m) \in \{0, 1\}^{N_{AR}(m)}$. Similarly, Node-B and Node-R generate $N_{BR}(m)$ secret bits, denoted by $k_{BR}(m) \in \{0, 1\}^{N_{BR}(m)}$, using their observations $\{y^{(2)}_{R}(l), 1 \leq l \leq L\}$ and $\{y^{(3)}_{B}(l), 1 \leq l \leq L\}$, respectively. The process of pair wise key generation involves an appropriate consensus algorithm to ensure that keys generated separately are identical. The specific consensus algorithm applied in our work would be discussed in Section \ref{sec:asym_throuput} and Section \ref{sec:prac_constraints}. 

\subsubsection{Broadcast Phase}

In the $(L+1)$-th coherence-block, Node-R intends to confidentially share $k_{AR}(m)$ with Node-B, using their secret-key $k_{BR}(m)$. Since the lengths of $k_{AR}(m)$ and $k_{BR}(m)$ are potentially different, the protocol for generating the broadcast message must be carefully designed. If $N_{AR}(m) < N_{BR}(m)$, then all the bits of $k_{AR}(m)$ can be confidentially shared with Node-B by XORing it with the corresponding number of bits in $k_{BR}(m)$. On the other hand, if $N_{AR}(m) > N_{BR}(m)$, then only $N_{AR}(m) - N_{BR}(m)$ bits of $k_{AR}$ can be shared due to lack of padding bits from $k_{BR}(m)$ to provide confidentiality. To circumvent this loss in key length, our buffer-aided protocol uses the pre-shared bits in the buffer $B(m-1)$ to generate a new sequence $k_{XOR}(m) \in \{0, 1\}^{N_{XOR}(m)}$ defined as in \eqref{eq:XOR_operation_generic},
\begin{figure*}
\begin{equation}
	\label{eq:XOR_operation_generic}
	k_{XOR}(m) = \left\{ \begin{array}{cccccccccc}
		k_{AR}(m) \oplus k_{BR}(m), & \mbox{ if } N_{AR}(m) = N_{BR}(m);\\
		k_{AR}(m) \oplus \bar{k}_{BR}(m), & \mbox{ if } N_{AR}(m) < N_{BR}(m);\\
		\bar{k}_{AR}(m) \oplus k_{BR}(m), & \mbox{ if } N_{AR}(m) - N_{BR}(m) > B(m-1);\\
		k_{AR}(m) \oplus \bar{\bar{k}}_{BR}(m), & \mbox{ if } N_{AR}(m) - N_{BR}(m) \le B(m-1);\\
	\end{array}
	\right.
\end{equation}
\hrule
\end{figure*}
where $B(0) = 0$, $\bar{k}_{BR}(m)$ constitutes the first $N_{AR}(m)$ components of $k_{BR}(m)$ and $\bar{k}_{AR}(m)$ constitutes the first $N_{BR}(m)$ components of $k_{AR}(m)$, and finally, $\bar{\bar{k}}_{BR}(m)$ constitutes the concatenation of $N_{BR}(m)$ and the additional $N_{AR}(m) - N_{BR}(m)$ bits from the buffer. With the use of the buffer it is clear that the length of the key is shortened only if sufficient number of padding bits is not available in the buffer (as captured in the third case of \eqref{eq:XOR_operation_generic}). Concurrently, in order to enhance the key length for the subsequent rounds, the buffer gets updated as in \eqref{eq:buffer_operation_generic},
\begin{figure*}
\begin{equation}
	\label{eq:buffer_operation_generic}
B(m) = \left\{ \begin{array}{cccccccccc}
		B(m-1), & \mbox{ if } N_{AR}(m) = N_{BR}(m);\\
		B(m-1) + N_{BR}(m) - N_{AR}(m), & \mbox{ if } N_{AR}(m) < N_{BR}(m);\\
		B(m-1), & \mbox{ if } N_{AR}(m) - N_{BR}(m) > B(m-1);\\
		B(m-1) - (N_{AR}(m) - N_{BR}(m)), & \mbox{ if } N_{AR}(m) - N_{BR}(m) \le B(m-1).\\
	\end{array}
	\right.
\end{equation}
\end{figure*}
\begin{figure*}
\begin{equation}
\label{recovery_node_B}
\hat{k}_{AR}(m) = \left\{ \begin{array}{cccccccccc}
\hat{k}_{XOR}(m) \oplus k_{BR}(m), & \mbox{ if } N_{XOR}(m) = N_{BR}(m);\\
\hat{k}_{XOR}(m) \oplus \bar{k}_{BR}(m), & \mbox{ if } N_{XOR}(m) < N_{BR}(m);\\
\hat{k}_{XOR}(m) \oplus [k_{BR}(m) ~u_{BR}(m)], & \mbox{otherwise}.\\
\end{array}
\right.
\end{equation}
\hrule
\end{figure*}
where $B(0) = 0$. Since $k_{AR}(m)$ and $k_{BR}(m)$ are uniformly distributed and statistically independent, the XOR operation in \eqref{eq:XOR_operation_generic} provides confidentiality to the underlying secret-key $k_{AR}(m)$ from an external eavesdropper \cite{lyw}, \cite{hlll}. As the last part of the broadcast phase, $k_{XOR}(m)$ is mapped to an $L$-length codeword $\mathbf{c} \in \mathcal{S} \subset \mathbb{C}^{L}$, and then sent to Node-B in the $(L+1)-$th coherence-block. Here, $\mathcal{S}$ denotes the chosen channel code of block-length $L$ in order to provide reliability. Assuming $\frac{1}{L}\mathbb{E}[|\mathbf{c}|^{2}] = 1$, Node-B receives
\begin{equation}
\label{eq:p4_eq1}
\mathbf{y}_{B} = \sqrt{(1-\beta) P} h_{BR}(L+1) \mathbf{c} + \mathbf{n}_{B} \in \mathbb{C}^{L},
\end{equation}
where $\mathbf{n}_{B}$ represents the AWGN, distributed as $\mathcal{CN}(\mathbf{0}_{L}, \gamma \mathbf{I}_{L})$. In addition to sending the codeword, Node-R also shares the length of $k_{XOR}(m)$ through a control channel. Then, Node-B decodes to $\hat{\mathbf{c}} \in \mathcal{S}$ using an appropriate decoder, and then recovers $\hat{k}_{XOR}(m)$. Using $N_{XOR}(m)$ (that is sent by Node-R) and $N_{BR}(m)$ (that is known during the pair-wise key generation process), the shared secret-key $\hat{k}_{AR}(m)$ is extracted as in \eqref{recovery_node_B}, where $\bar{k}_{BR}(m)$ constitutes the first $N_{AR}(m)$ components of $k_{BR}(m)$, and $u_{BR}(m)$ are the $N_{XOR}(m) - N_{BR}(m)$ bits retrieved from the buffer $B(m-1)$ at Node-B. The XOR operation in \eqref{recovery_node_B} can be viewed as successive interference cancellation (SIC) as $k_{BR}(m)$ and $u_{BR}(m)$ are perfectly known at Node-B. We highlight that $u_{BR}(m)$ was known since Node-R maintains an identical buffer as that at Node-B. After the recovery process, the buffer at Node-B also gets updated to obtain $B(m)$ identically at Node-R. Thus, due to the proposed buffer-aided relay, $N_{XOR}(m)$ bits are communicated to Node-B during the $L+1$ coherence-blocks of one round. However, due to the channel conditions in the $(L+1)$-th coherence-block, $\hat{k}_{XOR}(m)$ is likely to be in error despite using the channel code $\mathcal{S}$. As a result, we need to study the fraction of times $\hat{k}_{XOR}(m)$ is correctly decoded at Node-B. 

From the description of relay-assisted protocol, it is clear that the length of the secret-key that is correctly generated between Node-A and Node-B at the $m$-th round depends on the power-allocation factor $\beta \in (0,1)$, the LOS component $c_{R} \in [0, 1]$, the signal-to-noise ratio of the two channels, and the buffer-size $B(m-1)$ at the end of the $(m-1)$-th key generation round. To characterize the best-case benefits of the buffer-aided protocol on the $m$-th round, in the next section, we present the throughput and key rate analysis when optimal key generation algorithms are employed for pair-wise key generation between Node-A and Node-R, and Node-B and Node-R.
Throughout the next section, we drop the reference to $m$, when referring to the keys and their lengths.

\section{Performance Analysis with Optimal Key Generation Algorithms}
\label{sec:asym_throuput}

During the key generation phase of the protocol in Section \ref{buffer_protocol}, Node-A and Node-R collect the set $\{y^{(3)}_{A}(l), 1 \leq l \leq L\}$ and $\{y^{(1)}_{R}(l), 1 \leq l \leq L\}$, wherein $y^{(3)}_{A}(l)$ is correlated with $y^{(1)}_{R}(l)$ for each $l$ due to channel reciprocity. On this batch of samples, we assume that an optimal key generation algorithm \cite{nitin_narayan} is applied to obtain $k_{AR}$, wherein optimality is measured in terms of maximizing the average key-rate, which is $I\left(y^{(3)}_{A}(l); y^{(1)}_{R}(l)\right)$ bits per coherence-block. Note that the consensus methodology used is as per Section IV of \cite{nitin_narayan}. Similarly, Node-B and Node-R collect the set $\{y^{(3)}_{B}(l), 1 \leq l \leq L\}$ and $\{y^{(2)}_{R}(l), 1 \leq l \leq L\}$, and then apply an optimal key generation algorithm \cite{nitin_narayan} to obtain $k_{BR}$, with average key rate $I\left(y^{(3)}_{B}(l); y^{(2)}_{R}(l)\right)$ per coherence-block. Since we have considered same LOS parameter $c_R$ for both channels, i.e., channel between Node-A and Node-R, and channel between Node-B and Node-R, both the channels are identically distributed. With the same noise variances associated with the channels, the signals observed by all nodes are identically distributed. Since the optimal key rate is obtained by computing the mutual information of pair wise observations which provides identical results for identical distributions, key lengths generated between two pairs of nodes are identical, i.e., $N_{AR} = N_{BR}$.
After the XOR operation in \eqref{eq:XOR_operation_generic}, where the first condition is always satisfied, the sequence $k_{XOR}$ has to be broadcast to Node-B in $L$ channel-uses in the $(L+1)$-th coherence-block. Given that the entropy of the source is $I\left(y^{(3)}_{A}(l); y^{(1)}_{R}(l)\right)$ bits per coherence-block, and these bits are communicated over $L$ channel-uses, we represent the rate of communication in bits per channel-use for the broadcast phase as 
\begin{equation}
	\begin{split}
		M  \triangleq  \frac{1}{L}\mbox{log}_{2}\left(2^{N_{AR}}\right)
		&=   I\left(y^{(3)}_{A}(l); y^{(1)}_{R}(l)\right)\\
		\label{eq:mi_relay}
		&=  \mbox{log}_2 \left( 1 + \frac{(\frac{1 - c_{R}}{2})^2 \beta^2 \rho^2}{9+6(\frac{1 - c_{R}}{2})\beta \rho} \right),
	\end{split}
\end{equation}
where the last equality is computed using the joint distribution of $y^{(3)}_{A}(l)$ and $y^{(1)}_{R}(l)$ \cite{lyw}, \cite{waljin}. In \eqref{eq:mi_relay}, we define $\rho \triangleq \frac{P}{\gamma}$. From the joint source-channel coding theorem \cite{shannon_book}, it is well known that a source with entropy $I\left(y^{(3)}_{A}(l); y^{(1)}_{R}(l)\right)$ bits per sample cannot be reliably communicated over a channel with mutual information less than its entropy. Applying these results in our case, information-theoretic outage event will occur when the instantaneous mutual information of the channel between Node-B and Node-R in the $(L+1)$-th coherence-block is less than $M$. In other words, the probability that the channel $h_{BR}(L+1)$ is in outage is given by
\begin{equation}
	\label{eq:outage_prob}
	\resizebox{1\columnwidth}{!}{$
	\begin{split}
	P^{(BR)}_{out} & =  \mbox{Prob}\left(M \geq \mbox{log}_{2}\left(1 + |h_{BR}(L+1)|^2 (1-\beta)\rho\right)\right), \\
	& = \mbox{Prob}\left(|h_{BR}(L+1)|^2 \leq \left(\frac{1}{(1 - \beta)\rho}\right) \frac{(\frac{1 - c_{R}}{2})^2 \beta^2 \rho^2}{9+6(\frac{1 - c_{R}}{2})\beta \rho}\right),
	\end{split}$}
\end{equation}

where $\mbox{log}_{2}\left(1 + |h_{BR}(L+1)|^2 (1-\beta)\rho\right)$ is the mutual information of the channel from Node-R to Node-B. Since $|h_{BR}(L+1)|$ is Rician distributed, the corresponding outage-probability $P^{(BR)}_{out}$ given in \eqref{eq:outage_prob} can be computed in closed form. Specifically, using the cumulative distribution function (CDF) of a non-central Chi-Square random variable, we can write $\mbox{Prob}(|h_{BR}(L+1)|^2 \leq h)$, for $h \geq 0$, as
\begin{eqnarray}
\label{eq:outage_expr_Marcum_Q}
\resizebox{0.9\columnwidth}{!}{$
\mbox{Prob}\left(|h_{BR}(L+1)|^2 \leq h\right) = 1 - Q_{1}\left(\frac{\sqrt{c_{R}}}{\sqrt{\frac{1-c_{R}}{2}}},\frac{\sqrt{h}}{\sqrt{\frac{1-c_{R}}{2}}}\right)$},
\end{eqnarray}
where $Q_{1}(\cdot,\cdot)$ is a first-order Marcum-Q function expressed as
\begin{equation}
\label{eq:marcum_Q_function}
Q_{1}(\alpha,\lambda) = e^{-\frac{(\alpha^2 + \lambda^2)}{2}}\sum_{n=0}^\infty \left(\frac{\alpha}{\lambda}\right)^n I_n(\alpha \lambda),
\end{equation}
such that $I_n(\cdot)$ is the $n$-th order modified Bessel function. In the next section, we formally define a throughput metric, which captures the fraction of secret bits that are correctly recovered at Node-B.

\subsection{Throughput Analysis}
\label{subsec:throughput}

%
Over the $L+1$ coherence-blocks, Node-B recovers $LM$ secret bits through the relay channel whenever the channel from Node-R to Node-B is not in outage. On the other hand, Node-B recovers zero bits when the channel from Node-R to Node-B is in outage. Therefore, the average number of secret bits generated through the relay channel over $L+1$ coherence-blocks is $ML \left(1 - P^{(BR)}_{out}\right)$.

\begin{definition}
To capture the fraction of secret bits that reach Node-B from Node-R, we formally define the throughput of the scheme as
\begin{eqnarray}
\label{first_tp_expression}
\Theta \triangleq M \left(1 - P^{(BR)}_{out}\right),
\end{eqnarray}
where $M$ is given in \eqref{eq:mi_relay} and $P^{(BR)}_{out}$ is given in \eqref{eq:outage_prob}. Note that we have discarded $L$ in the throughput expression since the latency-interval is fixed. 
\end{definition}

In the above definition, we have assumed that $k_{XOR}$ can be accurately recovered at Node-B when $h_{BR}(L+1)$ is not in outage, assuming that the channel code $\mathcal{S}$ is appropriately designed. By using \eqref{eq:outage_expr_Marcum_Q} in \eqref{first_tp_expression}, we get
\begin{eqnarray}
\label{eq:tp_expression}
\Theta = M \times Q_{1}\left(\frac{\sqrt{c_{R}}}{\sqrt{\frac{1-c_{R}}{2}}},\frac{\sqrt{h}}{\sqrt{\frac{1-c_{R}}{2}}}\right),
\end{eqnarray}
where
\begin{equation}
\label{eq:h_expression}
h = \left(\frac{1}{(1 - \beta)\rho}\right) \frac{(\frac{1 - c_{R}}{2})^2 \beta^2 \rho^2}{9+6(\frac{1 - c_{R}}{2})\beta \rho}.
\end{equation}

Note that $\Theta$  is a function of $\beta, c_{R}$ and $\rho$. For a given $\rho > 0$ and a given $c_{R} \in [0, 1]$, our goal is to solve \footnote{We consider $\beta$ only for key generation and broadcast phases; consensus power is not considered in the optimization problem because as in lemma 4.3 of \cite{nitin_narayan} power required to achieve consensus is function of $\beta$ or directly proportional to power allocated for key generation phase.}

	\begin{equation}
		\label{eq:solve_TP}
		\displaystyle \mbox{$\beta$}_{opt}  = \max_{\beta \in (0, 1)} \Theta.
	\end{equation} 

	It is easy to verify that $\Theta \geq 0$ over the domain $\beta \in [0, 1]$ with equality when $\beta = 0$ and $\beta = 1$. This is because $M$ and $Q_{1}\left(\frac{\sqrt{c_{R}}}{\sqrt{\frac{1-c_{R}}{2}}},\frac{\sqrt{h}}{\sqrt{\frac{1-c_{R}}{2}}}\right)$ evaluate to zero when $\beta = 0$ and $\beta = 1$, respectively. Before solving \eqref{eq:solve_TP},
we need to understand the behavior of $\Theta$  as a function of $\beta$. We immediately note that $M$ is non-concave in $\beta$ and $Q_{1}\left(\frac{\sqrt{c_{R}}}{\sqrt{\frac{1-c_{R}}{2}}},\frac{\sqrt{h}}{\sqrt{\frac{1-c_{R}}{2}}}\right)$ is the sum of infinite Bessel functions. Although the Marcum-Q function has been shown to exhibit monotonous and log-concave properties \cite{SBZ_log_concave}, the function $Q_{1}\left(\frac{\sqrt{c_{R}}}{\sqrt{\frac{1-c_{R}}{2}}},\frac{\sqrt{h}}{\sqrt{\frac{1-c_{R}}{2}}}\right)$ is not log-concave in $\beta$ since the parameter $h$ in the second variable of the Marcum-Q function is a non-linear function of $\beta$, as shown in \eqref{eq:h_expression}. Furthermore, we also note that the Marcum-Q function is typically expressed as an infinite sum of the modified Bessel functions of the first kind. Thus, owing to the non-concave structure of $M$ and $Q_{1}\left(\frac{\sqrt{c_{R}}}{\sqrt{\frac{1-c_{R}}{2}}},\frac{\sqrt{h}}{\sqrt{\frac{1-c_{R}}{2}}}\right)$, we are unable to characterize the structure of their product as a function of $\beta$. As a result, the throughput expression is mathematically intractable to derive results on the optimal power-allocation factor.

To circumvent this issue, we present a lower bound on $\Theta$, and then analyze its behavior with respect to $\beta$. Subsequently, we will show that searching for $\beta$ that maximizes this lower bound on $\Theta$ will provide throughput values close to that when maximizing the exact throughput expression.


\subsection{Lower Bound on Throughput}
\label{subsec:lb_throughput}

In this section, we present a lower bound on $\Theta$, and then prove that the lower bound is unimodal over $\beta \in (0, 1)$ under some constraints on $c_{R}$ and $\rho$.
\begin{theorem}
The throughput expression given in \eqref{eq:tp_expression} is lower bounded by equation \eqref{eq:imp_lower_bound} shown at the top of next page.
\end{theorem}

\begin{figure*}[t]
	\begin{equation}
		\label{eq:imp_lower_bound}
		\Theta_{LB} = \left\{ \begin{array}{cccccccccc}
			\mbox{log}_2\left(1 + \frac{(\frac{1 - c_{R}}{2})^2\beta^2\rho^2}{18}\right)e^{-\frac{c_{R}}{2(\frac{1 - c_{R}}{2})}} e^{-\frac{\beta}{12(1-\beta)}}, & \mbox{ if } 0 < \beta \le \frac{9}{6(\frac{1 - c_{R}}{2})\rho};\\
			\mbox{log}_2\left(1 + \frac{(\frac{1 - c_{R}}{2})\beta \rho}{12}\right)e^{-\frac{c_{R}}{2(\frac{1 - c_{R}}{2})}} e^{-\frac{\beta}{12(1-\beta)}}, & \mbox{ if } \frac{9}{6(\frac{1 - c_{R}}{2})\rho} < \beta < 1.\\
		\end{array}
		\right.
	\end{equation}
	\noindent\rule{7.2in}{0.4pt}
\end{figure*}

\begin{IEEEproof}
In \eqref{eq:marcum_Q_function}, the infinite series of $I_n(\alpha \lambda)$ can be lower bounded by considering the first dominant term as 
$Q_{1}(\alpha,\lambda) > e^{-\frac{\alpha^2 + \lambda^2}{2}} I_0(\alpha \lambda) > e^{-\frac{\alpha^2 + \lambda^2}{2}},$
where the last inequality is applicable because of the bound $I_0(\alpha \lambda) > 1$ given in \cite{Luke}. Using the above lower bound on the Marcum-Q function in \eqref{eq:tp_expression}, and by substituting $\alpha = \sqrt{\frac{c_{R}}{\frac{1-c_{R}}{2}}}$ and $\lambda = \sqrt{\frac{2^{M}-1}{(\frac{1-c_{R}}{2})(1-\beta)\rho}}$, the throughput expression is lower bounded as
\begin{equation*}
\mbox{log}_2 \left( 1 + \frac{(\frac{1-c_{R}}{2})^2 \beta^2 \rho^2}{9+6(\frac{1-c_{R}}{2})\beta \rho} \right) e^{-\frac{c_{R}}{2(\frac{1-c_{R}}{2})}} e^{-\frac{(\frac{1-c_{R}}{2})\beta^2 \rho}{2(9+6(\frac{1-c_{R}}{2})\beta \rho)(1-\beta)}}.
\end{equation*}
Since $9 + 6 (\frac{1-c_{R}}{2}) \beta \rho$ is lower bounded by $6 (\frac{1-c_{R}}{2}) \beta \rho$, we further bound the third term in the above product to get
\begin{eqnarray}
\label{eq:lower_bound_TP}
\mbox{log}_2 \left( 1 + \frac{(\frac{1-c_{R}}{2})^2 \beta^2 \rho^2}{9+6(\frac{1-c_{R}}{2})\beta \rho} \right) e^{-\frac{c_{R}}{2(\frac{1-c_{R}}{2})}} e^{-\frac{\beta}{12(1-\beta)}}.
\end{eqnarray}
With this lower bound, we now split the domain $\beta \in (0, 1)$ into two parts, namely: (i) $0 < \beta \leq \frac{9}{6(\frac{1-c_{R}}{2}) \rho}$ and (ii) $\frac{9}{6(\frac{1-c_{R}}{2}) \rho} < \beta < 1$, provided the constants $\rho$ and $(1 - c_{R})$ are such that $\frac{9}{6(\frac{1-c_{R}}{2}) \rho} < 1$. In the latter case, when $\frac{9}{6(\frac{1-c_{R}}{2}) \rho} < \beta$, we have
\begin{eqnarray*}
	\mbox{log}_2\left(1 + \frac{(\frac{1-c_{R}}{2})^2 \beta^2 \rho^2}{9+6 (\frac{1-c_{R}}{2}) \beta \rho}\right) > \mbox{log}_2\left(1 + \frac{ (\frac{1-c_{R}}{2}) \beta \rho}{12}\right).
\end{eqnarray*}
By substituting the above bound in \eqref{eq:lower_bound_TP}, we get 
\begin{equation*}
\Theta > \mbox{log}_2\left(1 + \frac{ (\frac{1-c_{R}}{2}) \beta \rho}{12}\right) e^{-\frac{c_{R}}{2(\frac{1-c_{R}}{2})}} e^{-\frac{\beta}{12(1-\beta)}},
\end{equation*}
when $\frac{9}{6(\frac{1-c_{R}}{2}) \rho} < \beta < 1$. Similarly, when $0 < \beta \leq \frac{9}{6(\frac{1-c_{R}}{2}) \rho}$, we have the inequality
\begin{equation*}
\mbox{log}_2\left(1 + \frac{(\frac{1-c_{R}}{2})^2 \beta^2 \rho^2}{9+6(\frac{1-c_{R}}{2})\beta \rho}\right) \geq \mbox{log}_2\left(1 + \frac{(\frac{1-c_{R}}{2})^2 \beta^2 \rho^2}{18}\right).\\
\end{equation*}
By substituting the above bound in \eqref{eq:lower_bound_TP}, we get
	\begin{equation*}
	\Theta \geq \mbox{log}_2\left(1 + \frac{(\frac{1-c_{R}}{2})^2 \beta^2 \rho^2}{18}\right) e^{-\frac{c_{R}}{2(\frac{1-c_{R}}{2})}} e^{-\frac{\beta}{12(1-\beta)}},
	\end{equation*}
when $0 < \beta \leq \frac{9}{6(\frac{1-c_{R}}{2}) \rho}$. Combining the above two cases, we obtain the lower bound on the throughput given in \eqref{eq:imp_lower_bound}. This completes the proof.
\end{IEEEproof}

In the rest of this section, we present three lemmas to show that the proposed lower bound in \eqref{eq:imp_lower_bound} is unimodal in $\beta$. This result ensures that we can apply a gradient descent algorithm of suitable step-size to find $\beta$ that maximizes \eqref{eq:imp_lower_bound}, i.e., to solve the problem: 
\begin{equation}
\label{eq:soln_of_lower_bound}
\displaystyle \beta^{*} = \arg   \max_{\beta \in (0, 1)} \Theta_{LB}.
\end{equation}

Our approach to prove the unimodal property is to partition the interval $(0, 1)$ into three regions, namely: $\mathcal{R}_{1} = (0, \beta_{min}], \mathcal{R}_{2} = (\beta_{min}, \frac{23}{24}), \mbox{ and } \mathcal{R}_{3} = [\frac{23}{24}, 1)$, where $\beta_{min} = \frac{9}{6(\frac{1-c_{R}}{2})\rho}$ provided $c_{R}$ and $\rho$ satisfy an appropriate constraint. Subsequently, we prove that $\Theta_{LB}$  is 
(i). an increasing function in $\mathcal{R}_{1}$ (given in Lemma \ref{lemma1}), (ii). concave in $\mathcal{R}_{2}$ (given in Lemma \ref{lemma2}), and (iii). a decreasing function in $\mathcal{R}_{3}$ (given in Lemma \ref{lemma3}).

\begin{lemma}
\label{lemma1}
$\Theta_{LB}$  is an increasing function of $\beta$ in the interval $0 < \beta \le \frac{9}{6(\frac{1-c_{R}}{2})\rho}$ provided $(1-c_{R})$ and $\rho$ are such that $(\frac{1-c_{R}}{2})\rho > 1.862$.
\end{lemma}
\begin{IEEEproof}
When $0 < \beta \le \frac{9}{6(\frac{1-c_{R}}{2})\rho}$, $\Theta_{LB}$  is of the form  (from \eqref{eq:imp_lower_bound})
\begin{equation*}
\mbox{log}_2\left(1 + \frac{(\frac{1-c_{R}}{2})^2\beta^2\rho^2}{18}\right)e^{-\frac{c_{R}}{2(\frac{1-c_{R}}{2})}} e^{-\frac{\beta}{12(1-\beta)}}.
\end{equation*}
Since the second term in the above product does not contain $\beta$, we need to show that
\begin{equation*}
\frac{d}{d \beta} \left(\mbox{log}_2\left(1 + \frac{(\frac{1-c_{R}}{2})^2\beta^2\rho^2}{18}\right) e^{-\frac{\beta}{12(1-\beta)}}\right) > 0,
\end{equation*}
when $0 < \beta \le \frac{9}{6(\frac{1-c_{R}}{2})\rho}$. After differentiating the above and rearranging terms, we must prove the following equivalent inequality
\begin{equation}
	\label{last_inequality_bound}
		\begin{split}
			\frac{(\frac{1-c_{R}}{2})^2\beta \rho^2}{9}12(1-\beta)^2 > \mbox{log}\left(1+\frac{(\frac{1-c_{R}}{2})^2\beta^2 \rho^2}{18}\right)\times \\ \left(1+\frac{(\frac{1-c_{R}}{2})^2\beta^2 \rho^2}{18}\right).\\ 
		\end{split}		
\end{equation}

Applying the inequality $\mbox{log}(1 + x) \leq x$ on the first term of the right hand side (RHS) of \eqref{last_inequality_bound}, it suffices to show that
\begin{eqnarray*}
	\resizebox{1\columnwidth}{!}{$
\frac{(\frac{1-c_{R}}{2})^2\beta \rho^2}{9}12(1-\beta)^2 > \frac{(\frac{1-c_{R}}{2})^2\beta^2 \rho^2}{18} \times \left(1+\frac{(\frac{1-c_{R}}{2})^2\beta^2 \rho^2}{18}\right)$}.
\end{eqnarray*}
By again rearranging the above, we need to show
\begin{equation}
\label{eq:last_lower_bound}
24(1+\beta^2) - \frac{(\frac{1-c_{R}}{2})^2\rho^2\beta^3}{18} - 49\beta >  0,
\end{equation}
in the interval $0 < \beta \le \frac{9}{6(\frac{1-c_{R}}{2})\rho}$. We substitute $\beta = \frac{9}{6(\frac{1-c_{R}}{2})\rho}$ in the left hand side (LHS) of \eqref{eq:last_lower_bound}, and then find the range of values of $(\frac{1-c_{R}}{2})\rho$ such that the inequality in \eqref{eq:last_lower_bound} is satisfied. This implies we need to find the range of values of $(\frac{1-c_{R}}{2})\rho$ such that 
\begin{equation}
\label{eq:interim_inequality}
384\left(\left(\frac{1-c_{R}}{2}\right)\rho\right)^2 - 1179\left(\left(\frac{1-c_{R}}{2}\right)\rho\right) + 864 > 0.
\end{equation}
The roots of the above quadratic equation are $(\frac{1-c_{R}}{2})\rho = 1.21$ and $(\frac{1-c_{R}}{2})\rho = 1.862$. It is easy to note that the inequality in \eqref{eq:interim_inequality} is satisfied when $0 \le (\frac{1-c_{R}}{2})\rho < 1.21$ and $(\frac{1-c_{R}}{2})\rho > 1.862$. Out of the two regions, note that $(\frac{1-c_{R}}{2})\rho < 1.21$ leads to $\beta > 1$, and therefore this range is not applicable. On the other hand, the range $(\frac{1-c_{R}}{2})\rho > 1.862$ implies $\beta < 0.805$, and therefore this region is applicable. Thus, if $(\frac{1-c_{R}}{2})\rho > 1.862$, the extreme ends of the interval $\beta \in (0, \frac{9}{6(\frac{1-c_{R}}{2})\rho})$ satisfy \eqref{eq:last_lower_bound}. Finally, since the LHS of \eqref{eq:last_lower_bound} is a deceasing function of $\beta$, the inequality in \eqref{eq:last_lower_bound} is satisfied when $\beta$ takes interior points of the interval $(0, \frac{9}{6(\frac{1-c_{R}}{2})\rho})$. Thus, $\Theta_{LB}$  is an increasing function of $\beta$ in the interval $0 < \beta \le \beta_{min}$, where $\beta_{min} = \frac{9}{6(\frac{1-c_{R}}{2})\rho}$ provided $(\frac{1-c_{R}}{2})\rho > 1.862$.
\end{IEEEproof}

\begin{lemma}
\label{lemma2}
$\Theta_{LB}$  is a concave function when $\frac{9}{6(1-c_{R})\rho} < \beta < \frac{23}{24}$ provided $c_{R}$ and $\rho$ are such that $\frac{9}{6(1-c_{R})\rho} < \frac{23}{24}$.
\end{lemma}
\begin{IEEEproof}
From \eqref{eq:imp_lower_bound}, we have $\Theta_{LB} = \mbox{log}_2\left(1 + \frac{(\frac{1-c_{R}}{2})\beta \rho}{12}\right) e^{-\frac{c_{R}}{2(\frac{1-c_{R}}{2})}} e^{-\frac{\beta}{12(1-\beta)}}$.
Since the second term does not contain $\beta$, we represent $\mbox{log}_2\left(1 + \frac{(\frac{1-c_{R}}{2})\beta \rho}{12}\right)$ and $e^{-\frac{\beta}{12(1-\beta)}}$ as $f(\beta)$ and $t(\beta)$, respectively. Subsequently, we prove that the second derivative of $f(\beta)t(\beta)$ is negative, i.e., 
\begin{equation}
\label{eq:double_diff_lower_bound}
\frac{d^{2} f(\beta)}{d \beta^{2}} t(\beta) + 2 \frac{d f(\beta)}{d \beta} \frac{d t(\beta)}{d \beta} + f(\beta)\frac{d^{2} t(\beta)}{d \beta^{2}} < 0.
\end{equation}
Taking the first and the second derivatives of $t(\beta)$ and $f(\beta)$, we get
\begin{eqnarray*}
	\frac{dt\beta}{d\beta} & = & \frac{-t(\beta)}{12(1-\beta)^2},
	\frac{dt^2(\beta)}{\beta^2}  =  t(\beta)\frac{24\beta-23}{12^2(1-\beta)^4}.
\end{eqnarray*}
\begin{eqnarray*}
	\begin{split}
	\frac{df(\beta)}{d\beta} &=  \left(\mbox{log}_2e\right) \frac{1}{1+\frac{(\frac{1-c_{R}}{2})\beta\rho}{12}}\frac{(\frac{1-c_{R}}{2})\rho}{12},\\
	\frac{df^2(\beta)}{d\beta^2}  &=  -\left(\mbox{log}_2e\right)\left(\frac{\frac{(\frac{1-c_{R}}{2})\rho}{12}}{1+\frac{(\frac{1-c_{R}}{2})\beta\rho}{12}}\right)^2,
	\end{split}
\end{eqnarray*}

Since $\frac{df^2(\beta)}{d\beta^2}$ and $t(\beta)$ are always negative and positive, respectively, in the region of interest, the first term in \eqref{eq:double_diff_lower_bound} is always negative and so is the second term.
Since $\frac{d^{2} t(\beta)}{d \beta^{2}} < 0$ for $\beta < \frac{23}{24}$, and $f(\beta)$ is non-negative, it is straightforward to observe that \eqref{eq:double_diff_lower_bound} is negative in the region $\beta < \frac{23}{24}$. Furthermore, since the lower bound is applicable when $\beta > \frac{9}{6(\frac{1-c_{R}}{2})\rho}$, we deduce that $\Theta_{LB}$  is a concave function in the interval $\frac{9}{6(\frac{1-c_{R}}{2})\rho} < \beta < \frac{23}{24}$. This completes the proof.
\end{IEEEproof}

\begin{lemma}
\label{lemma3}
When $c_{R}$ and $\rho$ are such that $\frac{9}{6(\frac{1-c_{R}}{2})\rho} < \frac{23}{24}$, the lower bound on throughput given in \eqref{eq:imp_lower_bound} is a decreasing function of $\beta$ in the interval $\frac{23}{24} \le \beta < 1$.
\end{lemma}
\begin{IEEEproof}
The expression for $\Theta_{LB}$  is a product of three terms, where the second term is not a function of $\beta$. As a result, it suffices to prove that 
\begin{equation*}
\frac{d}{d \beta} \left(\mbox{log}_2\left(1 + \frac{(\frac{1-c_{R}}{2})\beta \rho}{12}\right)e^{-\frac{\beta}{12(1-\beta)}}\right) < 0,
\end{equation*}
in the interval $\frac{23}{24} < \beta < 1$. After differentiating the above with respect to $\beta$, we have to prove that

\begin{equation*}
	\begin{split}
		\mbox{log}_2\left(1+\frac{(\frac{1-c_{R}}{2})\beta \rho}{12}\right)\frac{-e^{-\frac{\beta}{12(1-\beta)}}}{12(1-\beta)^{2}} + \\ 
		(\mbox{log}_2{e}) \left(\frac{\frac{(\frac{1-c_{R}}{2})\rho}{12}}{1+\frac{(\frac{1-c_{R}}{2})\beta \rho}{12}}\right)e^{-\frac{\beta}{12(1-\beta)}} < 0.
	\end{split}
\end{equation*}

After rearranging the above terms, it suffices to prove that
\begin{equation*}
	\resizebox{1\columnwidth}{!}{$
\left(\frac{1-c_{R}}{2}\right)\rho(1-\beta)^2 < \left({1+\frac{(\frac{1-c_{R}}{2})\beta \rho}{12}}\right)\mbox{log}\left(1+\frac{(\frac{1-c_{R}}{2})\beta \rho}{12}\right)$}.	 
\end{equation*}



\noindent Applying the bound $\mbox{log}(x) \geq 1 - \frac{1}{x}$ on the RHS of the above equation, we get

\begin{equation*}
	\begin{split}
		\left({1+\frac{(\frac{1-c_{R}}{2})\beta \rho}{12}}\right)\mbox{log}\left(1+\frac{(\frac{1-c_{R}}{2})\beta \rho}{12}\right) \geq\\ \left({1+\frac{(\frac{1-c_{R}}{2})\beta \rho}{12}}\right)\left(\frac{\frac{(\frac{1-c_{R}}{2})\beta \rho}{12}}{1+\frac{(\frac{1-c_{R}}{2})\beta \rho}{12}}\right).
	\end{split}
\end{equation*}

As a result proving the below inequality suffices.
\begin{eqnarray*}
\left(\frac{1-c_{R}}{2}\right)\rho(1-\beta)^2 &<& {\frac{(\frac{1-c_{R}}{2})\beta \rho}{12}},\\
12(1-\beta)^2 &<& \beta.
\end{eqnarray*}
This further implies that we need to prove $12 - 25\beta + 12\beta^2 <  0.$ It can be verified that the roots of the quadratic equation $12 - 25\beta + 12\beta^2 = 0$ are $\frac{3}{4}$ and $\frac{4}{3}$. Therefore, $12 - 25\beta + 12\beta^2 > 0$ for $\beta < \frac{3}{4}$ and $\beta > \frac{4}{3}$, and also $12 - 25\beta + 12\beta^2 < 0$ for $\frac{3}{4} < \beta < \frac{4}{3}$. The interval $\frac{23}{24} \le \beta \le 1$ falls inside the interval $\frac{3}{4} < \beta < \frac{4}{3}$, and hence, the lower bound on throughput $\Theta_{LB}$  is a decreasing function of $\beta$ in the interval $\frac{23}{24} \le \beta < 1$.
\end{IEEEproof}

\begin{theorem}
\label{thm:unimodal}
$\Theta_{LB}$  given in \eqref{eq:imp_lower_bound} is unimodal provided $c_{R}$ and $\rho$ are such that $(\frac{1-c_{R}}{2})\rho > 1.862$.
\end{theorem}
\begin{IEEEproof}
The proof follows from the conjunction of results in Lemma \ref{lemma1}, Lemma \ref{lemma2} and Lemma \ref{lemma3}.
\end{IEEEproof}

	\begin{table*}[t]
		\centering
		\caption{\label{tab:beta_values}Comparison of optimized power-allocation parameters $\beta_{opt}$ and $\beta^{*}$, which are obtained by maximizing the exact expression of throughput in \eqref{first_tp_expression} and the lower bound in \eqref{eq:imp_lower_bound}, respectively.}
		\begin{tabular}{|l|c|c|c|c|c|c|c|c|c|c|c|c|}
			\hline
			$c_{R} \setminus \rho$ &\multicolumn{2}{|c|}{5dB} &\multicolumn{2}{|c|}{10dB} &\multicolumn{2}{|c|}{15dB} &\multicolumn{2}{|c|}{20dB}&\multicolumn{2}{|c|}{25dB}&\multicolumn{2}{|c|}{30dB}\\ \hline
			& $\beta_{opt}$ & $\beta^*$  & $\beta_{opt}$ & $\beta^*$ & $\beta_{opt}$ & $\beta^*$ & $\beta_{opt}$ & $\beta^*$ & $\beta_{opt}$ & $\beta^*$ & $\beta_{opt}$ & $\beta^*$ \\ \hline
			0   & 0.807 & 0.74  & 0.746 & 0.725 & 0.687 & 0.695 & 0.633 & 0.653 & 0.586 & 0.609 & 0.545 & 0.567 \\ \hline
			0.1 & 0.817 & 0.741 & 0.757 & 0.727 & 0.698 & 0.698 & 0.644 & 0.657 & 0.597 & 0.613 & 0.557 & 0.571 \\ \hline
			0.2 & 0.828 & 0.742 & 0.77  & 0.729 & 0.712 & 0.702 & 0.658 & 0.662 & 0.611 & 0.618 & 0.571 & 0.575 \\ \hline
			0.3 & 0.841 & 0.743 & 0.784 & 0.731 & 0.728 & 0.706 & 0.675 & 0.667 & 0.628 & 0.623 & 0.588 & 0.58 \\\hline
			0.4 & 0.856 & 0.79  & 0.802 & 0.733 & 0.747 & 0.71  & 0.696 & 0.673 & 0.649 & 0.629 & 0.61  & 0.585 \\\hline
			0.5 & 0.873 & 0.812 & 0.822 & 0.736 & 0.771 & 0.715 & 0.721 & 0.679 & 0.676 & 0.636 & 0.637 & 0.592 \\\hline
			0.6 & 0.894 & 0.813 & 0.848 & 0.738 & 0.8   & 0.72  & 0.753 & 0.687 & 0.711 & 0.644 & 0.674 & 0.6 \\\hline
			0.7 & 0.918 & 0.814 & 0.879 & 0.741 & 0.837 & 0.726 & 0.796 & 0.696 & 0.758 & 0.655 & 0.725 & 0.611 \\\hline
			0.8 & 0.948 & 0.815 & 0.918 & 0.75  & 0.885 & 0.733 & 0.854 & 0.708 & 0.824 & 0.671 & 0.798 & 0.627 \\\hline
			0.9 & 0.981 & 0.815 & 0.966 & 0.814 & 0.948 & 0.74  & 0.93  & 0.725 & 0.914 & 0.695 & 0.901 & 0.653 \\\hline
		\end{tabular}
	\end{table*}%
\subsection{Simulation Results on Throughput Optimization}
\label{subsec:sim_throughput}
In this section, we present simulation results on throughput optimization when the channel offered by Node-R experiences LOS values of $c_{R} \in \{0, 0.1, \ldots, 0.9\},$ and when the underlying SNR values are $\rho\in\{5,10,15,20,25,30\}$ in dB. For each combination of $c_{R}$ and $\rho$, the power-allocation parameter $\beta$ is obtained using the following methods: (i) Maximizing $\Theta$  in \eqref{eq:tp_expression} using a brute-force search over $\beta \in (0, 1)$ in steps of $0.001$, (ii) Applying a gradient descent method to maximize $\Theta_{LB}$  in \eqref{eq:imp_lower_bound} with step-size of $0.001$, (iii) Using $\beta = \frac{3}{4}$, which corresponds to equal power-allocation for key generation and broadcast phase. In Table \ref{tab:beta_values}, we list the values of $\beta_{opt}$ and $\beta^{*}$, which maximize the exact throughput expression and its lower bound, respectively. Table \ref{tab:beta_values} highlights that the absolute values of $\beta$ offered by solving the optimization problem are different from that of uniform distribution, i.e., $\beta = \frac{3}{4}$. 

In Fig. \ref{fig:TP_benefits}, we present the plots on $\Theta$ , which are obtained by substituting $\beta$ from (i), (ii) and (iii). 
The plots highlight that at 
 $ \rho = 15$ dB, the throughput values of all the three schemes are approximately same. However, at $\rho = 20, 25, \mbox{ and }30$ dB, the plots show that instead of allocating equal power to key generation and broadcast phase, the choice of $\beta$ must be made by solving the proposed optimization problem as a function of $c_{R}$ and $\rho$. In particular, Fig. \ref{fig:TP_benefits} indicates that the gains over equal power-allocation is maximum when $\rho$ is high and $c_{R} = 0$, which corresponds to Rayleigh channel between Node-R and Node-B. Furthermore, the plots show that the throughput values obtained by maximizing the lower bound is approximately same as that when optimizing the exact throughput expression.
\subsection{Key Rate Analysis}
\label{sec:key_rate}

In this section, we consider the metric of maximizing the key rate subject to an upper bound on the outage probability of the broadcast phase, given by $P^{(BR)}_{out} \leq \eta$, for some $\eta > 0$. 
 The key rate maximization problem for a given $\eta > 0$ can be formally defined as below,
 \begin{eqnarray}
 	\label{eq:opt_constraint_1}
 	& \displaystyle \arg \max_{\beta \in (0, 1)} \mbox{log}_2 \left( 1 + \frac{(\frac{1-c_{R}}{2})^2 \beta^2 \rho^2}{9+6(\frac{1-c_{R}}{2})\beta \rho} \right),\\
 	& \mbox{ such that }P^{(BR)}_{out} \leq \eta. \nonumber
 \end{eqnarray}
We consider the following lemmas to solve the optimization problem in \eqref{eq:opt_constraint_1}.
\begin{lemma}
	\label{lemma_increasing_rate}
	The key rate in \eqref{eq:mi_relay} is an increasing function of $\beta$.
\end{lemma}
\begin{IEEEproof}
	Taking the first derivative of equation \eqref{eq:mi_relay} w.r.t $\beta$, we get
	\begin{eqnarray*}
		\frac{d}{d \beta}M &=& \frac{d}{d\beta}\mbox{log}_2 \left( 1 + \frac{(\frac{1 - c_{R}}{2})^2 \beta^2 \rho^2}{9+6(\frac{1 - c_{R}}{2})\beta \rho} \right),\\
		&=& \mbox{log}_2 e \frac{1}{9+6(\frac{1 - c_{R}}{2})\beta \rho + (\frac{1 - c_{R}}{2})^2 \beta^2 \rho^2}\\
		&\times& \frac{18(\frac{1 - c_{R}}{2})^2\beta\rho + 12(\frac{1 - c_{R}}{2})^3\beta^2\rho^3}{9+6(\frac{1 - c_{R}}{2})\beta\rho}.
	\end{eqnarray*}
For $c_R \in [0,1]$, $\beta \in (0,1)$, and $\rho > 0$, $\frac{d}{d \beta}M$ is always positive, and thus the key rate, M is an increasing function of $\beta$.
\end{IEEEproof}
\begin{lemma}
	\label{lemma_increasing_outage}
	The outage probability in \eqref{eq:outage_expr_Marcum_Q} is an increasing function of $\beta$.
\end{lemma}
\begin{IEEEproof}
	Taking the first derivative of equation \eqref{eq:outage_expr_Marcum_Q} w.r.t $\beta$, gives
	\begin{eqnarray*}
		\frac{d}{d \beta}P_{out}^{(BR)} &=& -\frac{d}{d\beta}Q_{1}\left(\frac{\sqrt{c_{R}}}{\sqrt{\frac{1-c_{R}}{2}}},\frac{\sqrt{h}}{\sqrt{\frac{1-c_{R}}{2}}}\right)\\
		&=& \frac{1}{1-c_R}I_o\left(\frac{2\sqrt{c_Rh}}{1-c_R}\right)e^{-\frac{c_R + h}{(1-c_R)}}\times\\
		&& \frac{(\frac{1-c_{R}}{2})^2\beta\rho\left(18-9\beta+6(\frac{1-c_{R}}{2})\beta\rho\right)}{\left(9+6(\frac{1-c_{R}}{2})\rho\beta-9\beta-6(\frac{1-c_{R}}{2})\beta^2\rho\right)^2}.\\
	\end{eqnarray*}
In the RHS on the above equation, the term $I_o(\frac{2\sqrt{c_Rh}}{1-c_R})>1$ as given in \cite{Luke}. The denominator of $\frac{d}{d \beta}P_{out}^{(BR)}$ is always positive. Furthermore, for $\beta \in (0,1)$, the numerator of the last product term is also always positive. Thus, $P_{out}^{(BR)}$ is an increasing function of $\beta$.
\end{IEEEproof}
With the above two lemmas, the optimization problem in \eqref{eq:opt_constraint_1} can be  transformed into an equivalent problem, as given in the following theorem.

\begin{theorem}
\label{thm:optimal_beta}
The optimization problem in \eqref{eq:opt_constraint_1} has same optimal solution as that of the following problem.	
\end{theorem}
\begin{eqnarray}
		\label{eq:opt_constraint_new}
		& \displaystyle \arg \max_{\beta \in (0, 1)} \mbox{log}_2 \left( 1 + \frac{(\frac{1-c_{R}}{2})^2 \beta^2 \rho^2}{9+6(\frac{1-c_{R}}{2})\beta \rho} \right),\\
		& \mbox{ such that }P^{(BR)}_{out} = \eta. \nonumber
\end{eqnarray}


\begin{IEEEproof}
Suppose that $\beta^{\dagger}$ is the solution that satisfies the outage probability equation in \eqref{eq:opt_constraint_new} for a given value of $\eta$. Since the outage probability is an increasing function of $\beta$ by Lemma \ref{lemma_increasing_outage}, the solutions for the outage probability inequality equation in \eqref{eq:opt_constraint_1} for the same value of $\eta$ will be in the interval $(0,\beta^{\dagger}]$. Furthermore, since the key rate expression (which is the objective function of \eqref{eq:opt_constraint_1} and \eqref{eq:opt_constraint_new}) is an increasing function of $\beta$ by Lemma \ref{lemma_increasing_rate}, the optimal solution for the optimization problem in \eqref{eq:opt_constraint_1} will also be $\beta^{\dagger}$. Thus, the optimal solution to \eqref{eq:opt_constraint_1} and \eqref{eq:opt_constraint_new} are the same.
\end{IEEEproof}

\begin{figure*}[t]
	\begin{minipage}[c]{0.95\columnwidth}
		\includegraphics[scale=0.362]{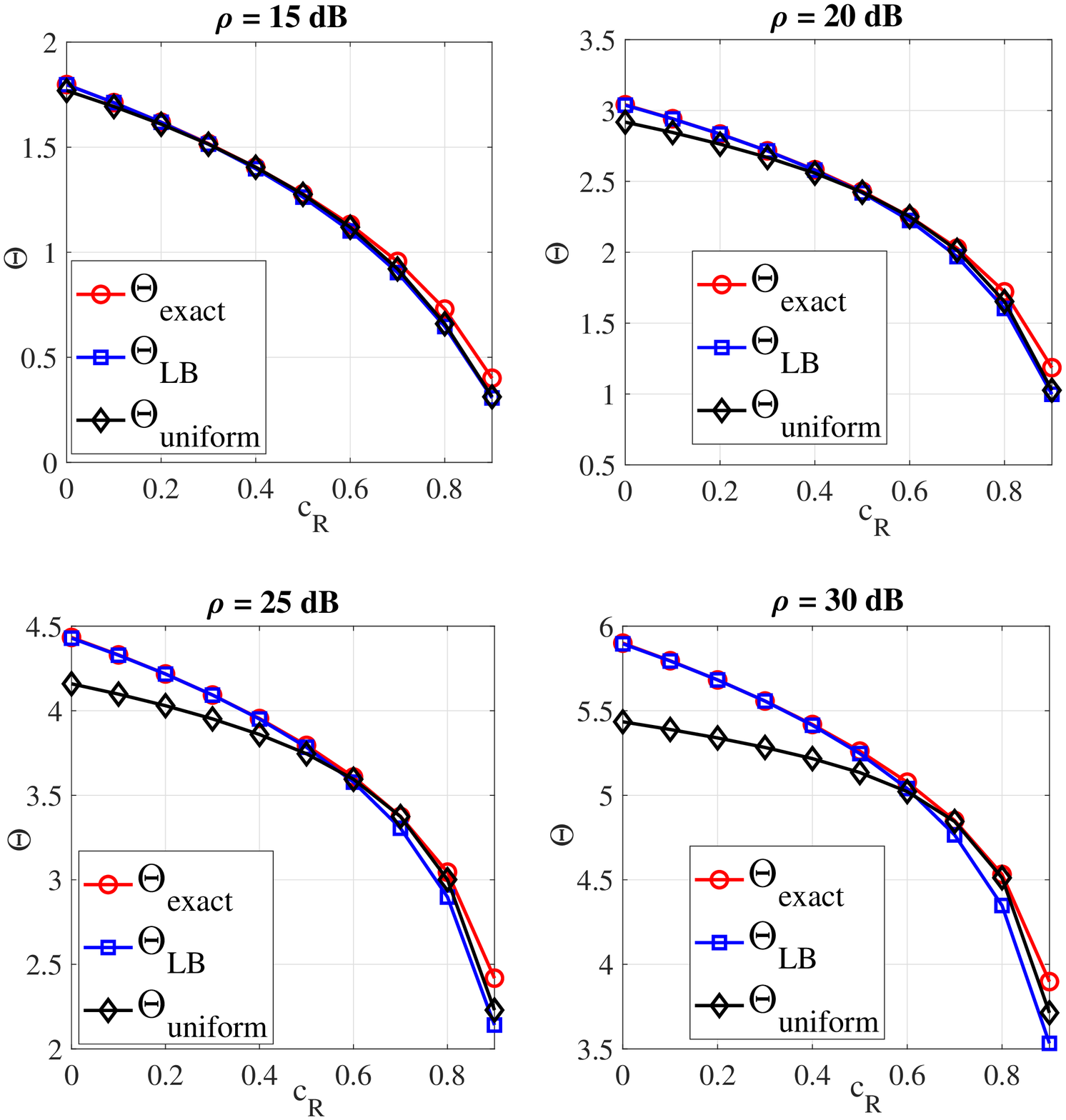}
	\end{minipage}
	\begin{minipage}[c]{0.8\columnwidth}
		\centering
		\includegraphics[scale=0.38]{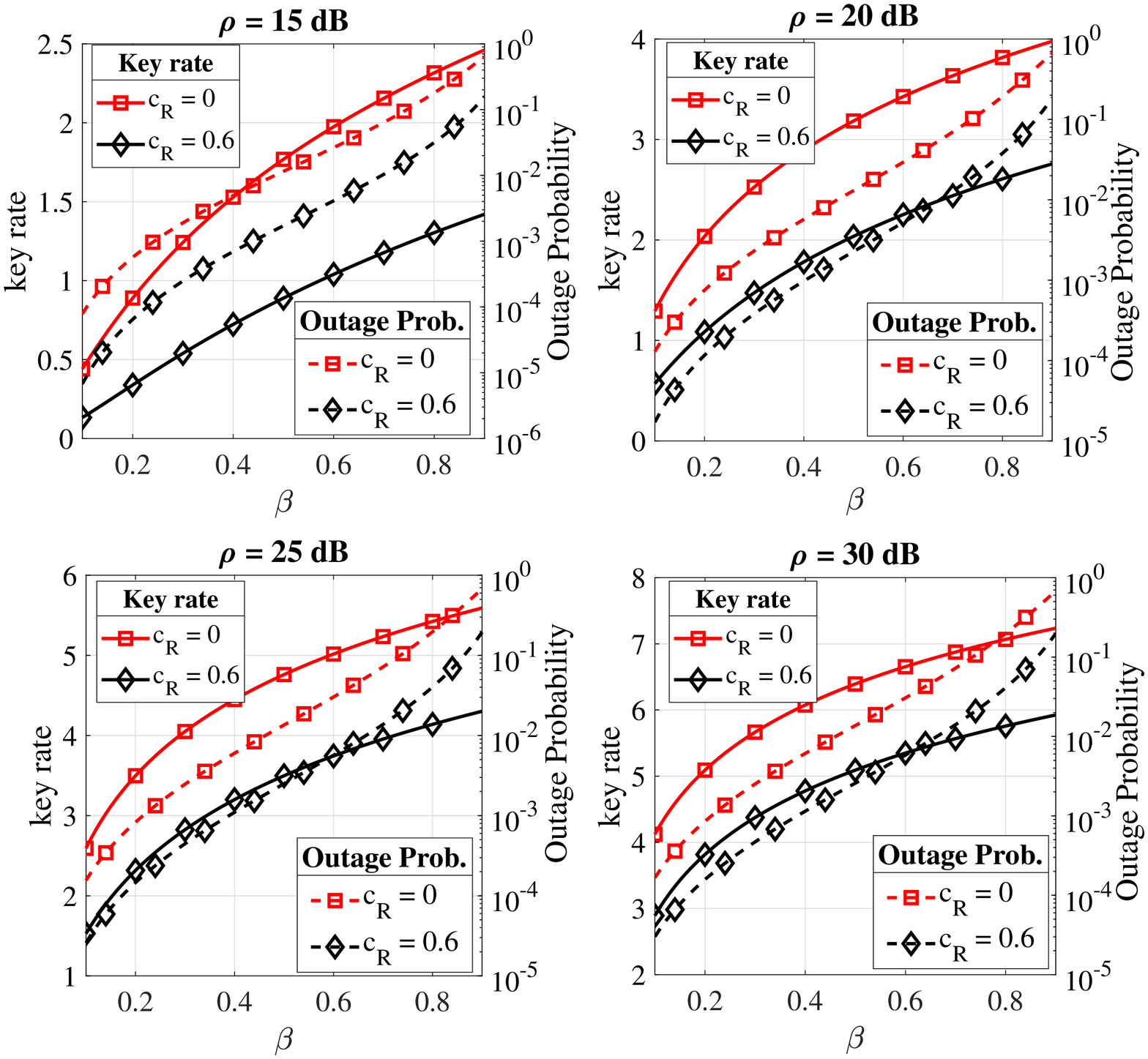}
	\end{minipage}
	\vspace{-0.5cm}
	\caption{\label{fig:TP_benefits} \textbf{On the left:} Plots of throughput when $\beta$ is optimized using (i) $\Theta_{\mbox{exact}}$, the exact throughput expression in \eqref{first_tp_expression}, (ii) the lower bound $\Theta_{LB}$  given in \eqref{eq:imp_lower_bound}, and (iii) $\Theta_{\mbox{uniform}}$, which corresponds to equal power-allocation for the key generation phase and the broadcast phase. \label{fig:key_rate} \textbf{On the right:} Plots of key rate and its outage probability as a function of $\beta$ with various values of $\rho$ and $c_{R}$, while considering that block length is asymptotically large.}
\end{figure*}

Note that the outage probability, $P_{out}^{(BR)}$ is the complement of the first order Marcum Q-function, which is mathematically intractable. Thus, solving the root of the equality in \eqref{eq:opt_constraint_new} for a given $\eta$, requires a numerical approach. As a naive approach to numerically solve this problem, we need to plot the key rate along with the corresponding outage probability expression as a function of $\beta$ for a given $\rho$ and $c_R$ (as shown in Fig. \ref{fig:TP_benefits}). Subsequently, to determine the maximum key rate with the outage probability constraint, we should draw a horizontal line $P^{(BR)}_{out} = \eta$ on the plots and then determine the point at which this line intersects the outage probability curve. The corresponding x-coordinate of this point of intersection gives the $\beta$ value that maximizes the key rate under the constraint. However, as a formal approach to solve \eqref{eq:opt_constraint_new}, a possible strategy is to use the well-known Newton-Raphson method \cite{NR}, wherein the function $P^{(BR)}_{out} - \eta$ is used to iteratively compute the root by choosing a suitable step-size for $\beta$ in the range $(0, 1)$.

%


\section{Performance Analysis with Practical Key Generation Algorithms}
\label{sec:prac_constraints}
	
In this section, we discuss the buffer-aided protocol of Section \ref{sec:relay_model} from an implementation viewpoint by considering practical key generation algorithms at the three nodes, and an empty buffer at the start of the key generation protocol. 
\subsection{Buffer-aided Relay Protocol}

In this section, we adopt the relay-assisted key generation protocol discussed in Section \ref{buffer_protocol}, with the exception that practical key generation algorithms are used to optimize the throughput and the key rate yard-sticks. Adopting the state-of-the art key generation protocol \cite{swnc_practical_key} on their observations, Node-A and Node-R, unfold each complex number into two real values. Subsequently, a two-level crossing algorithm is applied on the samples by choosing an appropriate guard band, denoted by $q^{-}$ and $q^{+}$, such that the key miss-match rate is less than or equal to $\epsilon$, for some $\epsilon > 0$. After applying the algorithm in \cite{swnc_practical_key}, Node-R generates a key $k_{AR}$ of length $N_{AR}$ with Node-A, and also generates a key $k_{BR}$ of length $N_{BR}$ along with Node-B. Since $N_{AR}$ and $N_{BR}$ are random variables, we are interested in characterizing  the average key lengths $\mathbb{E}[N_{AR}]$ and $\mathbb{E}[N_{BR}]$. Towards that direction, the following lemma is straightforward to prove. 

\begin{lemma}
\label{lemma4}
With $y_{A}$ and $y_{R}$ denoting the real sample of a coherence-block between Node-A and Node-R, the probability that the two samples are in consensus is $p_{\epsilon}= \iint\limits_{y_A,y_R \notin (q-, q+)} f(y_A,y_R)dy_Ady_R,$ where $f(y_A,y_R)$ denotes the joint density function of $y_{A}$ and $y_{R}$, and $q+$ and $q-$ represent the threshold levels of the guard band of the two-level crossing algorithm. 
\end{lemma}

The consensus probability for samples between Node-B and Node-R is also $p_{\epsilon}$ since the two channels are identically distributed. With $2L$ real samples subject to the two-level key generation algorithm, and since each real sample is statistically identical and independent, the key length between a pair of nodes is equal to the number of real samples in consensus. Thus, we note that both $N_{AR}$ and $N_{BR}$ are independent Binomial random variables defined as $N_{AR} \sim \mathcal{B}in(2L,p_{\epsilon})$, and $N_{BR} \sim \mathcal{B}in(2L,p_{\epsilon})$, and therefore, we have  $\mathbb{E}[N_{AR}] = \mathbb{E}[N_{BR}] = 2Lp_{\epsilon}$. In case, a multi-level crossing algorithm is used for key generation \cite{JJR_group_key}, the average key length will be multiplied by $f$, where $2^f$ is the number of levels of the quantizer. From the buffer-aided relay model, the length of $k_{XOR}$ depends on the difference between $N_{AR}$ and $N_{BR}$, and the size of the buffer at that point. In practice, the buffer may not have sufficient bits, and therefore, we expect $N_{XOR}$ to be less than $N_{AR}$ in some rounds of the key generation protocol. In the next section, we revisit the broadcast phase of the relay model in order to compute $\mathbb{E}[N_{XOR}]$ with buffer constraints.
\subsection{Broadcast Phase with Buffer Constraints}

\begin{figure*}[t]
	\begin{center}
	\hspace*{-2cm}
		\includegraphics[height = 5.5 cm, width = 22cm]{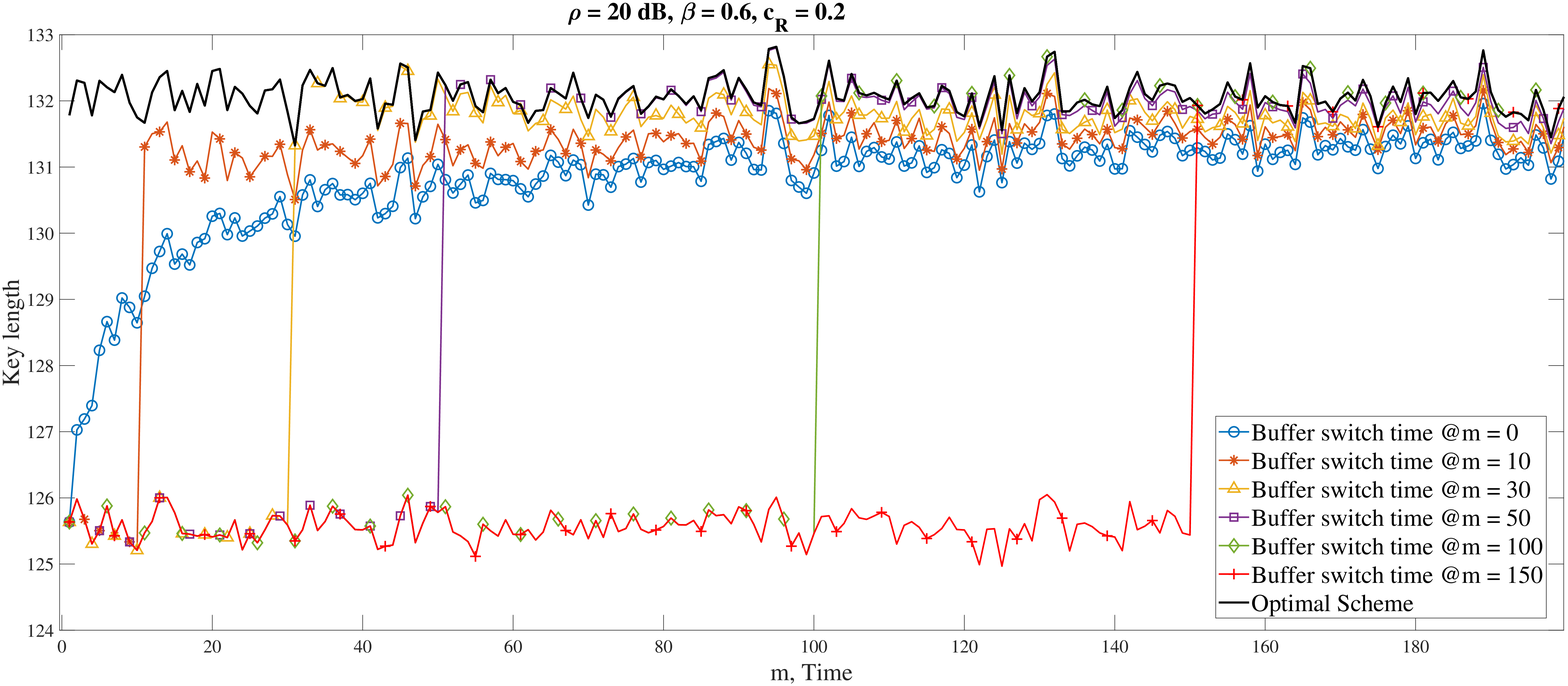}
	\end{center}
	\vspace{-0.5cm}
	\linespread{1}\selectfont{}
	\caption{\label{fig:finite_buffer} Plots of average key lengths as a function of time for SNR  $\rho = 20$ dB, power optimization parameter, $\beta = 0.6$, LOS parameter, $c_R = 0.2$, and number of coherence-blocks, $L = 100$ for variable bits available in the buffer at different instants of time.}
\end{figure*}

We recollect from Section \ref{buffer_protocol} that the buffer bits at Node-R are the unused bits in consensus between Node-B and Node-R from the previous \emph{rounds} of the key generation protocol. To incorporate practical constraints on the buffer for analyzing the throughput and key rate, we incorporate the round number $m \geq 0$ when referring to the buffer size, keys and their lengths. Consequently, $B(m), N_{AR}(m), N_{BR}(m)$, $N_{XOR}(m)$ denote the buffer size, the pair-wise key lengths, and the length of the sequence $k_{XOR}(m)$ broadcast to Node-B at the end of the $m$-th round. With these definitions, we immediately note that $N_{AR}(m), N_{BR}(m)$ are statistically independent across $m$, whereas $B(m)$ and $N_{XOR}(m)$ are statistically dependent across $m$ because the latter numbers depend on $B(m-1)$, $N_{AR}(m)$ and $N_{BR}(m)$. In particular, with the updates in \eqref{eq:XOR_operation_generic} and \eqref{eq:buffer_operation_generic}, we broadly have three types of schemes, namely: 
\begin{itemize}
\item the optimal scheme, wherein $B(m) = \infty$, $\forall m$, 
\item the min-scheme, wherein $B(m) = 0, ~\forall m$, and 
\item the intermediate scheme, where $B(0) = 0$ for $m = 0$.
\end{itemize}
We note that the optimal scheme is such that $\mathbb{E}[N_{XOR}(m)] = \mathbb{E}[N_{AR}(m)] = 2Lp_{\epsilon}$, for all $m$, whereas the min-scheme is such that $\mathbb{E}[N_{XOR}(m)] = \mathbb{E}[\mbox{min}(N_{AR}(m), N_{BR}(m))]$ for all $m$. Given that the intermediate scheme makes use of the buffer as and when available, it is intuitive that $\mathbb{E}[N_{XOR}(m)]$ of the intermediate scheme should lie in between that of the optimal scheme and the min-scheme. Moreover, unlike the optimal scheme and the min-scheme, we expect that $\mathbb{E}[N_{XOR}(m)]$ for the intermediate scheme changes as a function of $m$ since $\{B(m)~|~ m \geq 0\}$ is a non-stationary random process. Towards understanding the behavior of $\mathbb{E}[N_{XOR}(m)]$ of the three schemes, we present simulation results to compute the average key lengths of their broadcast phase, i.e., $\mathbb{E}[N_{XOR}(m)]$ for various values of $m$. In Fig. \ref{fig:finite_buffer}, we plot the average key lengths of all the three schemes as a function of $m$ for the channel condition $\rho = 20$ dB, $\beta = 0.6$ and $c_{R} = 0.2$. Furthermore, under the intermediate scheme, we plot the average key lengths for various buffer switch-on time-instants. In this context, the buffer switch-on time is the round index of the key generation protocol until which the min-scheme is active, i.e., the buffer is allowed to grow from $m = 0$ till the switch-on instant whenever $N_{AR}(m) < N_{BR}(m)$. To generate the simulation results, the buffer switch-on time were $\{0,10,30,50,100,150\}$. From the simulation results, we observe that the intermediate scheme achieves the  optimal scheme when the buffer is switched-on at $ \{100, 150\}$, and it is apparent that we do not need to wait longer than $m = 100$ to achieve the optimal key length.
Inspired by the observations made in Fig. \ref{fig:finite_buffer}, we would like to theoretically analyze the key lengths, i.e., $\mathbb{E}[N_{XOR}(m)]$ for any given $m$, offered by the three variants. Note that the average key lengths of these regimes are different from that of the asymptotic case because these average key lengths are from using a practical key generation algorithm as opposed to the optimal one as discussed in Section \ref{sec:asym_throuput}. In order to characterize $\mathbb{E}[N_{XOR}(m)]$, we need to understand the Probability Mass Function (PMF) on the difference between $N_{AR}(m)$ and $N_{BR}(m)$ (which are individually binomially distributed and statistically independent), and also the PMF on $B(m)$. In the following lemma, we present the PMF on $D(m) \triangleq N_{AB}(m) - N_{BR}(m)$ by dropping the index $m$ since $D(m)$ is identically distributed across $m$. 

\begin{lemma}
\label{lemma5}
Given $N_{AR}, N_{BR}  \sim \mathcal{B}in(2L,p_{\epsilon})$ and are statistically independent, the PMF on $D = N_{AR} - N_{BR}$, i.e., $P(D = d)$, for $-2L\leq d \leq 2L$, can be computed as in \eqref{eq:pmf_D}.
\end{lemma}

\begin{figure*}[t]
	\begin{equation}
		\label{eq:pmf_D}
			\begin{aligned}
			P(D = d) =  \left\{ \begin{array}{lllllll}
				\sum_{t = 0}^{2L + d} \binom{2L}{t}p^t(1-p)^{(2L-t)} \binom{2L}{t-d}p^{(t-d)}(1-p)^{(2L-t+d)}, &\mbox{if} -2L \le d < 0;\\
				\sum_{t = 0}^{2L - d} \binom{2L}{t+d}p^{(t+d)}(1-p)^{(2L-t-d)} \binom{2L}{t}p^t(1-p)^{(2L-t)},&\mbox{if   }   0 \le d \le 2L.\\
			\end{array}    
			\right.
			\end{aligned}
		\end{equation}
%
	\setcounter{equation}{25}
\begin{equation}
	\label{eq:pmf_B_m'+1}
	\begin{aligned}
		P(B(m'+1) = b_1) = \left\{ \begin{array}{cccccccccc}
			P(D(m'+1) > B(m')) + P(D(m'+1) = 0), &\mbox{if } b_1 = B(m');\\
			P(D(m'+1) = B(m') - b_1),&\mbox{if   }   0 \le b_1 \le B(m') + n$ \& $b_1 \neq B(m');\\
		\end{array}    
		\right.
	\end{aligned}
\end{equation}
%
\begin{equation}
	\label{eq:buffer_eqn}
	B(m'+j) = \left\{ \begin{array}{cccc}
		B(m'+ j -1) - D(m'+j) & \mbox{ if } D(m'+j) \le B(m'+j-1);\\
		B(m'+j-1), & \mbox{ if } D(m'+j) > B(m'+j-1);\\
	\end{array}    
	\right.
\end{equation}
	\noindent\rule{7.2in}{0.4pt}
\end{figure*}

In the following lemma, we provide a way to construct the PMF on $B(m)$ for any $m \geq 1$.
\begin{lemma}
\label{lemma6}
Assuming that the buffer is switched-on at round $m'$, the PMF on $B(m)$ can be computed in closed-form for any $m > m'$. \end{lemma}
\begin{IEEEproof}
Throughout the proof, we use $n = 2L$. Let $B(m') = B_{0} < n$ be the size of buffer at the time when buffer is switched on. At the $(m' + 1)^{th}$ round of key generation, we have
	\setcounter{equation}{24}
	\begin{equation}
		\resizebox{\columnwidth}{!}{$
		B(m' + 1) = \left\{ \begin{array}{cccc}
			B(m') - D(m'+1), & \mbox{ if } D(m' + 1) \le B(m');\\
			B(m'), & \mbox{ if } D(m' + 1) > B(m');\\
		\end{array}   
		\right.$} 
	\end{equation}
where $B(m' + 1) = N_{AR}(m'+ 1) - N_{BR}(m'+ 1)$ is a random variable which has a support set, $\mathcal{S}upp(B(m' + 1)) = [0, B(m') + n]$. Since $B(m')$ is a constant, the PMF of $B(m'+1)$ can be easily determined using the PMF of $D(m'+ 1)$ as in \eqref{eq:pmf_B_m'+1},
\noindent where $b_{1}$ is a realization of $B(m'+ 1)$. At the $(m'+j)^{th}$ round of key generation, for $j \geq 2$, the buffer size is updated using the distribution on $B(m'+ j-1)$ as given in \eqref{eq:buffer_eqn},
where $\mathcal{S}upp(B(m'+j)) = [0, B(m') + jn]$. Defining a new random variable $Y(m'+j) = B(m'+j-1) - D(m'+j)$ whose PMF can be determined, the PMF of $B(m'+j)$ can be determined as below by using $b_{j}$ as a realization of $B(m'+ j)$.\\
Case 1: $0 \le b_j \le B(m') + (j-1)n$:
\setcounter{equation}{27}
\begin{equation}
	\begin{aligned}
		&P(B(m'+j) =  b_j)\\ &= P(B(m'+j-1) = b_j)P(D(m'+j) > b_j) \\
		&+ P(Y(m'+j) = b_j).\\
	\end{aligned}
\end{equation}

\noindent Case 2: $B(m') + (j-1)n + 1 \le b_j \le B(m') + jn$:
\begin{equation}
	\begin{aligned}
		P(B(m'+j) =  b_j) &= P(Y(m'+j) = b_j).\\
	\end{aligned}
\end{equation}
This completes the proof.
\end{IEEEproof}	
Using the PMFs on $D(m)$ and $B(m)$, in the following theorem, we provide a closed form expression on the average key length of the intermediate that has $B(m-1) = b_{m-1}$ bits in the buffer for the $m$-th round.
%
%

%
\begin{figure*}[t]
	\setcounter{equation}{29}
	\begin{equation}
		\label{eq:average_keylength}
		\begin{aligned}
			&\mathbb{E}[N_{XOR}(m) ~|~ B(m-1)= b_{m-1}] \\
			&= \sum_{z_1 = b_{m-1} + 1}^n\Big \{\sum_{z_2 = 0}^{z_1 - b_{m-1} - 1}z_2P_{N_{BR}(m)}(z_2)\Big \}P_{N_{AR}(m)}(z_1)   + \sum_{z_2 = 0}^{n - b_{m-1}}\Big \{\sum_{z_1 = 0}^{b_{m-1} + z_2}z_1P_{N_{AR}(m)}(z_1)\Big \}P_{N_{BR}(m)}(z_2),
		\end{aligned}
	\end{equation}
	\setcounter{equation}{32}
	\begin{equation}
		\label{eq:average_keylength_mth_round}
		\mathbb{E}[N_{XOR}(m)] = \sum_{b_{m-1} \in \mathbb{S}upp(B(m-1))} P(B(m-1) = b_{m-1}) \mathbb{E}[N_{XOR}(m)| B(m-1) = b_{m-1}].
	\end{equation}
	\setcounter{equation}{34}
	\begin{equation}
		\label{eq:average_keylength_min}
		\begin{aligned}
			\mathbb{E}[N_{XOR}(m)]
			= \sum_{z_1 = 1}^n\Big \{\sum_{z_2 = 0}^{z_1 - 1}z_2P_{N_{BR}(m)}(z_2)\Big \}P_{N_{AR}(m)}(z_1)   + \sum_{z_2 = 0}^{n }\Big \{\sum_{z_1 = 0}^{z_2}z_1P_{N_{AR}(m)}(z_1)\Big \}P_{N_{BR}(m)}(z_2).
		\end{aligned}
	\end{equation}
	\noindent\rule{7.2in}{0.4pt}
	
\end{figure*}
\begin{theorem}
Given that $B(m-1) = b_{m-1}$, for some fixed $b_{m-1} > 0$, the average key length of the buffer-aided practical key generation scheme for the $m$-th round is given by \eqref{eq:average_keylength}, where $P_{N_{AR}(m)}(\cdot)$ and $P_{N_{BR}(m)}(\cdot)$ denote the evaluation of PMF of $N_{AR}(m)$ and $N_{BR}(m)$ at a particular realization, respectively.

	
\end{theorem}

\begin{IEEEproof}
By definition, $N_{XOR}(m)$ is given by
	\setcounter{equation}{30}
	\begin{equation}
		\label{eq:Z_defn}
		\resizebox{\columnwidth}{!}{$
		N_{XOR}(m) = \left\{ \begin{array}{cccc}
			N_{AR}(m), & \mbox{ if } N_{AR}(m) - N_{BR}(m) \le b_{m-1};\\
			N_{BR}(m), & \mbox{ if } N_{AR}(m) - N_{BR}(m) > b_{m-1}.\\
		\end{array}
		\right.$}
	\end{equation}
Since $b_{m-1}$ is a constant, we can use the joint probability mass function of $N_{AR}(m)$ and $N_{BR}(m)$, denoted by $P_{N_{AR}(m),N_{BR}(m)}(\cdot,\cdot)$, to compute $\mathbb{E}[N_{XOR}(m) ~|~ B(m-1) = b_{m-1}]$ as below:
\begin{equation}
\begin{aligned}
			&\mathbb{E}[N_{XOR}(m)/B(m-1) = b_{m-1}]\\ 
			&=\sum_{z_2}\sum_{z_1}zP_{N_{AR}(m),N_{BR}(m)}(z_1,z_2),\\
			&=\sum_{z_2}\sum_{z_1}zP_{N_{AR}(m)}(z_1)P_{N_{BR}(m)}(z_2),\\
			&=\sum_{z_2 = 0}^{n - b_{m-1}}\Big \{\sum_{z_1 = 0}^{b_{m-1} + z_2}z_1P_{N_{AR}(m)}(z_1)\Big \}P_{N_{BR}(m)}(z_2)\\ &+ \sum_{z_1 = b_k + 1}^n\Big \{\sum_{z_2 = 0}^{z_1 - b_k - 1}z_2P_{N_{BR}(m)}(z_2)\Big \}P_{N_{AR}(m)}(z_1),
\end{aligned}
\label{eq:case_1}
\end{equation}
wherein $z_{1}$ and $z_{2}$ run through different realizations of $N_{AR}(m)$ and $N_{BR}(m)$, while $z$ takes the appropriate value depending on $z_{1}, z_{2}$ and $b_{m-1}$, as defined in \eqref{eq:Z_defn}. This completes the proof.
\end{IEEEproof}

Finally, using the above result, the overall average key length for the $m$-th round can be computed as in \eqref{eq:average_keylength_mth_round}.
As a special case of the above expression, we can also obtain average key length of the optimal scheme and min-scheme. In particular, to obtain the optimal scheme as a special case, applying $B(m) = \infty, \forall   m$ in \eqref{eq:XOR_operation_generic}, we observe that $N_{AR}(m) - N_{BR}(m) > B(m-1)$ never happens, and therefore, we have
\setcounter{equation}{33}
\begin{eqnarray}
	\label{eq:infinite_averagekeylength}
	\mathbb{E}[N_{XOR}(m)] = \mathbb{E}[N_{AR}(m)]
		= \sum_{z_1 = 0}^{n}z_1P_{N_{AR}(m)}(z_1).
\end{eqnarray}
Note that the above expression corresponds to the black curve in Fig. \ref{fig:finite_buffer}. Similarly, we can also obtain the average key length of the min-scheme by using $B(m) = 0, \forall m$. As a result, we get \eqref{eq:average_keylength_min}.
We now need to use $\mathbb{E}[N_{XOR}(m)]$, for $m \geq 1$, to compute the power-allocation parameter $\beta$ such that the throughput is maximized or the key rate is maximized subject to the outage probability constraint in the broadcast phase.
\subsection{Throughput and Key Rate Analysis}

Based on our protocol for the $m$-th round, the key generation phase is executed with $\beta \in (0, 1)$, after which Node-R has an average of $\mathbb{E}[N_{XOR}(m)]$ secret bits from the first $L$ coherence-blocks. In the $(L+1)$-th coherence-block, Node-R communicates this message using a code of block-length $L$ with rate $R =\frac{\mathbb{E}[N_{XOR}(m)]}{L}$, such that the average power per channel use is $(1 - \beta)P$. Towards studying the error performance of the code used for the broadcast phase, the outage probability expression of Section \ref{sec:asym_throuput} is no longer applicable since $L$ need not be large in practice. As a result, we apply the corresponding non-asymptotic outage probability results of \cite{yuri_FBL} to compute $P^{(BR)}_{out}(m)$ as a function of $\beta$. Formally, $P^{(BR)}_{out}(m)$ can be computed as
\setcounter{equation}{35}
\begin{equation}
	P^{(BR)}_{out}(m) = \int_\mathbb{R} Q\left (  \sqrt{\frac{L}{V(\Gamma)}}(C(\Gamma)-R)\right )f_\Gamma(\Gamma)d\Gamma,
\end{equation}
where $\Gamma = |h_{BR}|^2(1 - \beta)\frac{P}{\gamma}$ is the instantaneous SNR of the wireless channel, $(1-\beta)P$ is the average power per channel use, $R$ is the code rate, $f_\Gamma(.)$ is the probability density of the instantaneous SNR $\Gamma$, $C(\Gamma) = log_2(1 + \Gamma)$ is the Shannon channel capacity, $V(\Gamma) = \frac{\Gamma}{2}\frac{\Gamma + 2}{(\Gamma+1)^2}log^2_2e$ is the back-off factor for finite block-length, and finally, $Q(x)= \frac{1}{2\pi}\int_{x}^{\infty}e^{-\frac{u^2}{2}}du$. When using a practical key generation algorithm along with finite block length code, we observe that both the average key length $\mathbb{E}[N_{XOR}(m)]$ and the outage probability $P^{(BR)}_{out}(m)$ are not available in closed-form, and therefore, we solve the optimization problem $\arg \max_{\beta \in (0, 1)} \mathbb{E}[N_{XOR}(m)](1 - P^{(BR)}_{out}(m))$ using simulation results.\footnote{Consensus power is not considered in the optimization problem as in the similar case of asymptotic analysis.}

\begin{figure*}[t]
	\begin{minipage}[c]{0.9\columnwidth}
		\includegraphics[scale = 0.36]{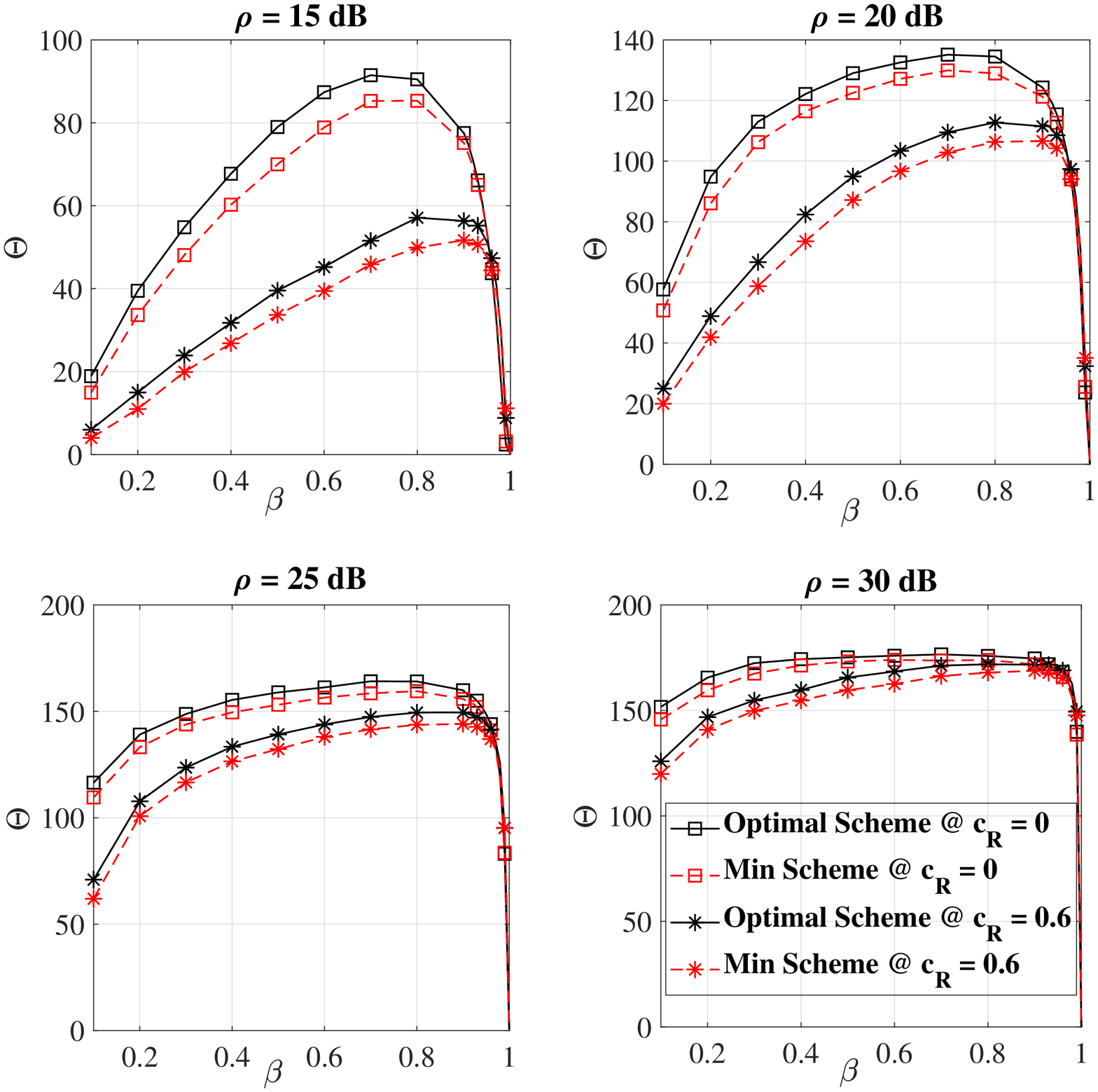}
	\end{minipage}
	\hfill{}
	\begin{minipage}[l]{1.04\columnwidth}
		\includegraphics[scale = 0.36]{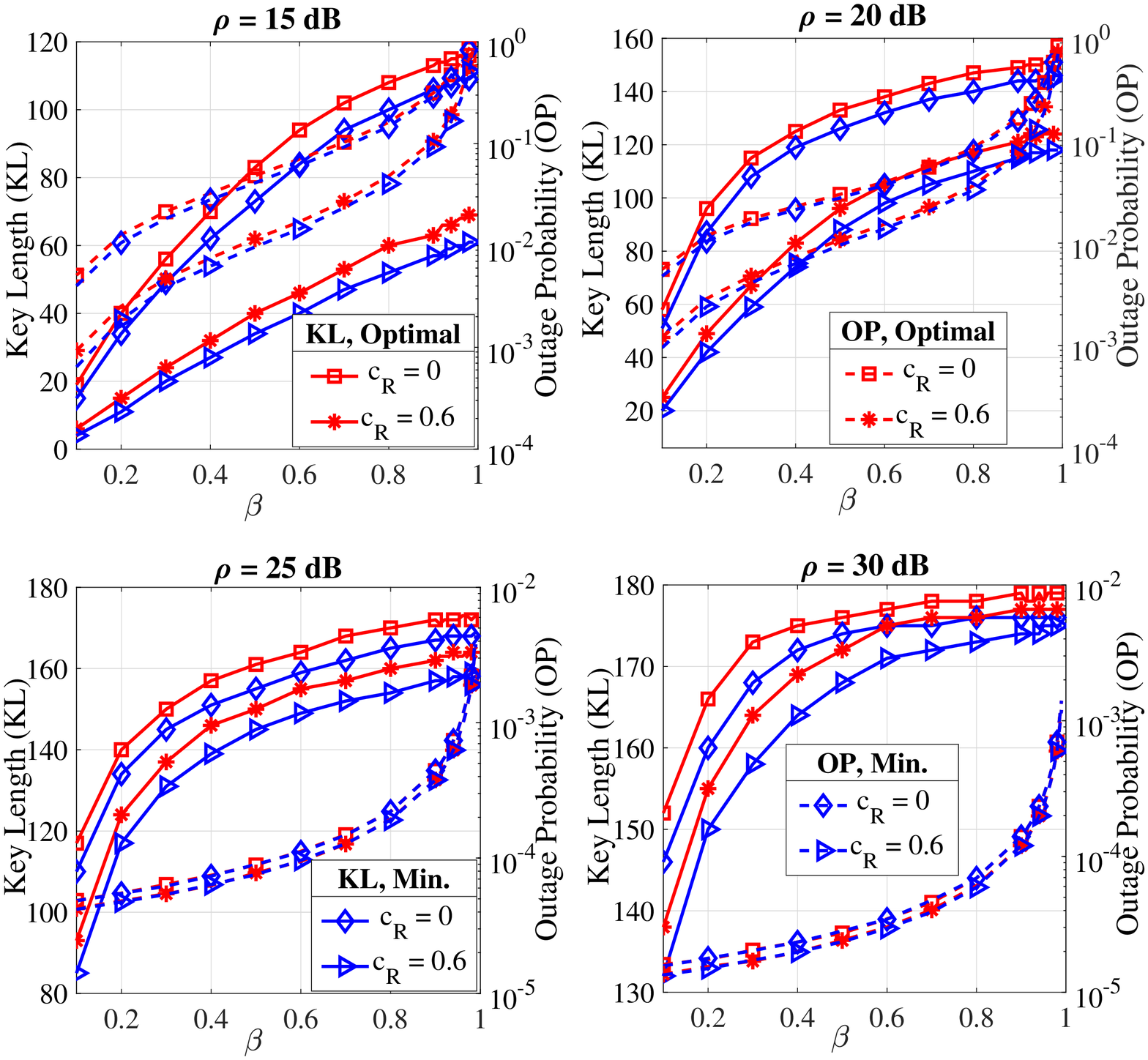}
	\end{minipage}
	\vspace{-0.5cm}
	\caption{\label{fig:TP_non} \textbf{On the left:} Plots of throughput as a function of $\beta$ with various values of $\rho$ and $c_{R}$. The legend in the fourth subplot applies to all subplots of left side. \label{fig:key_rate_non} \textbf{On the right:} Plots of average key length and its outage probability as a function of $\beta$ with various values of $\rho$ and $c_{R}$. Note that four legends in each subplot apply to all subplots.}
\end{figure*}

To generate the simulation results, we fix $L = 100$, and then vary the value of $\beta \in (0, 1)$ at discrete steps to compute $\mathbb{E}[N_{XOR}(m)](1 - P^{(BR)}_{out}(m))$ for both the optimal scheme and the min-scheme. These simulations are presented in Fig. \ref{fig:TP_non} for the LOS parameters $c_R = [0,0.6]$ and SNR, $\rho = [15, 20, 25, 30]$ in dB. We observe that the optimal value of $\beta$ that maximizes the throughput is close to $\beta = 0.9$, and this behavior is unlike the throughput analysis for the asymptotic case. This difference is because of two factors: first, in the broadcast phase the generated key is encoded using a strong channel code of block-length $L$, and second, the rate of increase of $\mathbb{E}[N_{XOR}(m)]$ as $\beta$ increases is low due to two-level quantizer employed in key generation process. In general, when either a multi-level crossing algorithm is used for key generation, or when the codes used for the broadcast is uncoded, we expect the optimal value of $\beta$ to reduce. We also observe from the plots that the optimal scheme outperforms the min-scheme in throughput \cite{lyw},\cite{hlll},\cite{qkk},\cite{waljin}. For instance, in Fig. \ref{fig:TP_non} when $\rho = 15$ dB and $c_R = 0$, optimal scheme with optimal power allocation yields $8\%$ increase in throughput, in comparison to min-scheme with equal power allocation.

%

Similar to the key rate analysis for the asymptotic case in Section \ref{sec:key_rate}, in Fig. \ref{fig:key_rate_non}, we plot the average key length and outage probability for the case of short block-lengths. We observe that both the average key length and the outage probability increase as $\beta$ increases. To determine the optimal value of $\beta$ that provides maximum key rate subject to given outage probability, we suggest drawing a horizontal line which corresponds to the given outage probability constraint, and then the x-coordinate of the intersection point of the line gives the optimal $\beta$ value. The plots in  Fig. \ref{fig:key_rate_non} show that for a given outage probability constraint, the optimal scheme provides higher key rate compared to the min-scheme, which is none other than the existing buffer-less key generation method \cite{lyw},\cite{hlll},\cite{qkk},\cite{waljin}. This behavior can be attributed to the fact that although the optimal scheme provides higher values of $\mathbb{E}[N_{XOR}(m)]$, the corresponding difference in the rate does not result in significant changes in the outage probability. For instance, in Fig. \ref{fig:key_rate_non} when $\rho = 15$ dB, $c_R = 0$, optimal scheme  yields $9\%$ increase in average key length, in comparison to min-scheme subject to outage probability,  $P^{(BR)}_{out} = 10^{-2}$. Overall, the simulation results of this section have shown that the optimal scheme with the help of the buffer outperforms the min-scheme \cite{lyw},\cite{hlll},\cite{qkk},\cite{waljin}. 

\section{Buffer-aided Protocol with Asymmetric LOS Characteristics}
\label{sec:unequal_c}

In Section \ref{sec:relay_model}, we introduced a network model wherein the LOS parameters of the channels between Node-A and Node-R, and Node-B and Node-R are identical (denoted by $c_{R} \in [0, 1]$). As a result, without loss of generality, the key generated between Node-A and Node-R during the key generation phase is broadcast to Node-B, although the roles of Node-A and Node-B could have been swapped. However, if a relay network is such that the channels between Node-A (or Node-B) and Node-R have different LOS parameters, then appropriate modifications on the buffer-aided protocol must be discussed. Towards that direction, the following proposition states that higher the LOS parameter of the channel, lower is the key rate offered by the PLK generation.

\begin{proposition}
	\label{prop:MI_decr_c}
	The mutual information $I(y_{A}^{(3)}(l);y_{R}^{(1)}(l))$ in \eqref{eq:mi_relay} is a decreasing function of $c_{R} \in [0, 1]$.
\end{proposition}

To formally discuss the modifications needed on the buffer-aided protocol, let $c_{AR} \in [0, 1]$ and $c_{BR} \in [0, 1]$ represent the LOS components of $h_{AR}$ and $h_{BR}$, respectively. Furthermore, suppose that $c_{AR} < c_{BR}$. During the key generation phase of the protocol, it is clear from Proposition \ref{prop:MI_decr_c} that the pair-wise keys $k_{AR}$ and $k_{BR}$ are such that $N_{BR} < N_{AR}$ with high probability in each round of the protocol. As a result, even if the relay assisted key generation is executed over several rounds to fill the buffer with unused bits of $k_{BR}$, there will not be sufficient bits to provide confidentiality to achieve $N_{AR}$ as the key length. Thus, choosing $k_{BR}$ as the key for broadcast is the best strategy, and hence, buffers are no longer useful. Note that the key $k_{BR}$ should not be broadcast to Node-A through $h_{AR}$ because transmitting a message over lower LOS channel results in higher values of outage probability compared to transmitting over a channel with higher LOS parameter. Because of this result, we propose that Node-R uses the first $N_{BR}$ bits of $k_{AR}$, and then broadcast an XOR version to Node-B. This way, both the key-rate of the key generation phase as well as the outage probability of the broadcast phase are functions of the LOS parameter $c_{BR}$. Consequently, the optimization problems for throughput and key rate for this case can be solved by applying the techniques  in Section \ref{sec:asym_throuput} and Section \ref{sec:prac_constraints}, however, by replacing $c_R$ with $c_{BR}$. Overall, we have shown that the power-allocation strategies proposed in this work are also applicable to a relay network when the LOS components of the two channels are different.
\section{Summary}
\label{sec:summary}

We addressed a key generation scenario wherein two wireless devices seek the assistance of a trusted relay to generate secret-keys. To handle the reduction in key rate due to XOR based broadcast, we have proposed buffer-aided key generation protocol at the relay wherein the unused secret bits generated between the relay and one of the nodes can be temporarily stored in a buffer before using them to provide the confidentiality feature for the broadcast phase in the subsequent rounds. We have also proposed power allocation policies between the key generation phase and broadcast phase by optimizing the overall key rate and throughput in different scenarios: (i) optimal or practical key generation methods, and (ii) empty buffer at the start of algorithm. Simulation results show that using buffer results in remarkable benefits when the LOS of the two channels are close.

Towards implementing the strategies proposed in this work, the knowledge of the LOS component is required for choosing an appropriate value of the power-allocation factor. In practice, we believe that this could be achieved in the following way. Before a pair of nodes initiates the key generation process, each node processes to detect the LOS component by estimating the fading channel characteristics. For instance, the techniques mentioned in \cite{Ref1, Ref2} could be used since LOS component is usually a long-term characteristic of the channel. Subsequently, for every round of secret key generation, each node could use the corresponding optimal power allocation parameter (solved using our proposed scheme) against the detected LOS component. One way to implement this in practice is to store a table of the power allocation parameter values and the LOS values a priori in the memory of the devices. With this stored information, in the subsequent rounds of secret key generation and distribution, the three nodes could allocate the power for their transmission by looking up the table in the memory.


\end{document}